\documentclass[lettersize,journal]{IEEEtran}

\usepackage{amsmath,amsfonts}
\usepackage{algorithmic}
\usepackage{algorithm}
\usepackage{array}
\usepackage[caption=false,font=normalsize,labelfont=sf,textfont=sf]{subfig}
\usepackage{textcomp}
\usepackage{stfloats}
\usepackage{url}
\usepackage{verbatim}
\usepackage{graphicx}
\usepackage{cite}
\usepackage{balance}
\usepackage{multirow}
\usepackage{subcaption}
\usepackage{xcolor}
\usepackage{etoolbox} 
\usepackage{ulem}
\AtBeginEnvironment{equation}{\small}
\AtBeginEnvironment{equation*}{\small}
\AtBeginEnvironment{align}{\small}
\AtBeginEnvironment{align*}{\small}

\def\BibTeX{{\rm B\kern-.05em{\sc i\kern-.025em b}\kern-.08em
    T\kern-.1667em\lower.7ex\hbox{E}\kern-.125emX}}

\newcommand{\new}[1]{{\color{black}#1}}

\begin{document}
\bstctlcite{IEEEexample:BSTcontrol}

\title{PoTAcc: A Pipeline for End-to-End Acceleration \\ of Power-of-Two Quantized DNNs}

\author{Rappy Saha,~\IEEEmembership{Member,~IEEE}, Jude Haris,~\IEEEmembership{Member,~IEEE},
        Nicolas Bohm Agostini,~\IEEEmembership{Member,~IEEE}, \\ David Kaeli,~\IEEEmembership{Fellow,~IEEE}, and José Cano,~\IEEEmembership{Senior Member,~IEEE}

\thanks{This work was partially supported by the UK Engineering and Physical Sciences Research Council (grants EP/T517896/1 and EP/W524359/1), the EU Project dAIEDGE (GA Nr 101120726) and the Innovate UK Horizon Europe Guarantee (GA Nr 10090788).}}




\maketitle


\begin{abstract}

Power-of-two (PoT) quantization significantly reduces the size of Deep Neural Networks (DNNs) and allows the use of bit-shifts as core operations instead of multiplications for DNN inference.
Previous studies have shown that PoT-quantized DNNs can maintain the accuracy of a target application, such as image classification, but it remains unclear how these DNNs perform when deployed on resource-constrained edge devices.
While general-purpose edge CPUs or GPUs lack optimized bit-shift kernels as backends for PoT-quantized DNN inference, custom hardware accelerators can leverage the benefits of PoT quantization by designing a bit-shift accelerator specific to a PoT quantization method. 
However, challenges arise because custom hardware accelerators often lack support from inference frameworks that enable the deployment of PoT-quantized DNNs.
Additionally, the effect of different PoT quantization methods on custom accelerator designs and their implications on performance and energy efficiency during full inference requires further exploration to enable efficient deployment of PoT-quantized DNNs on resource-constrained edge devices.

To bridge this gap, we propose \texttt{PoTAcc}, an open-source pipeline for efficient end-to-end acceleration and evaluation of PoT-quantized DNNs on resource-constrained edge devices.
Using \texttt{PoTAcc}, we demonstrate how to prepare and deploy PoT-quantized models via TensorFlow Lite (TFLite) on different hardware devices, including CPU-only and those with a CPU and a custom hardware accelerator.
We design custom hardware accelerators using efficient shift-based processing elements (shift-PE) for three different PoT quantization methods, while targeting two different FPGA platforms.
We evaluate the accuracy, performance, energy efficiency and resource utilization of these accelerators across a wide range of DNNs, including CNNs and Transformer-based models.
Our proposed accelerator paired with a CPU achieves up to $3.6\times$ speedup and a $78.0\%$ energy reduction when compared to CPU-only execution for different PoT-quantized DNNs on the PYNQ-Z2 and Kria boards.
Our code will be available at \texttt{https://github.com/gicLAB/PoTAcc}

\end{abstract}



\section{Introduction}
\label{sec:introduction}

\IEEEPARstart{D}{eploying} Deep Neural Networks (DNNs) on resource-constrained edge devices requires achieving a critical balance between efficiency and accuracy~\cite{gibsonDLASConceptualModel2025}. 
Ideally, the goal is to run DNN models on edge devices with floating-point accuracy while meeting performance and energy budgets.
One commonly used technique for achieving this is quantization.
While it is well known that 8-bit uniform quantization can achieve accuracy comparable to floating-point, previous research~\cite{li_additive_2020,chang_mix_2021,przewlocka-rus_power--two_2022,chang_mix_2021,xia_energy-and-area-efficient_2023} has shown that 4-bit power-of-two (PoT) quantization, a type of non-uniform quantization, can better match to a data distribution as compared to uniform quantization, achieving good accuracy using fewer bits. 
Additionally, PoT quantization allows replacing multiplications by simpler bit-shift operations.

In a resource-constrained edge device where reducing power consumption is crucial and design area is limited, PoT quantization can be a promising solution for running DNNs efficiently.
Understanding the benefits of PoT quantization for DNN models requires end-to-end deployment on the target hardware.
Since general-purpose hardware, such as CPUs and GPUs, is not optimized for bit-shift-based DNN inference, custom hardware accelerators are needed to fully leverage the benefits of PoT quantization.
However, running a PoT-quantized model on custom hardware requires an inference framework that supports custom hardware accelerators, as well as PoT-quantized models.
Existing inference frameworks, such as TensorFlow Lite~\cite{tflite} (or TFLite, now renamed as LiteRT), TensorRT~\cite{tensorrt} and TVM~\cite{chen2018tvm} provide support for custom hardware accelerators, but they do not support PoT-quantized models out of the box.

Previous work has proposed various PoT quantization methods to improve the accuracy of a DNN for different application domains~\cite{miyashita_convolutional_2016,zhou_incremental_2017,gudovskiy_shiftcnn_2017,li_additive_2020,elhoushi_deepshift_2021,chang_mix_2021,przewlocka-rus_power--two_2022,xia_energy-and-area-efficient_2023,jiang_jumping_2023,li2023densshift}.
However, only a few studies have explored the hardware implementations of their proposed PoT quantization methods~\cite{vogel_bit-shift-based_2019,liu_optimize_2020,chang_mix_2021,przewlocka-rus_power--two_2022,xia_energy-and-area-efficient_2023}.
While previous work has demonstrated the potential of PoT quantization methods, they have not investigated how these methods behave across different DNNs in the context of end-to-end deployment.
Here, end-to-end deployment refers to running a PoT quantized model on an accelerator where bit-shift operations replace multiplications.
The challenges come from two main aspects: (i) preparing the PoT-quantized DNN model that is trained on frameworks such as TensorFlow or PyTorch for the inference framework like TFLite; and (ii) designing an efficient accelerator that is integrated with inference frameworks and can leverage the computational benefits of PoT quantization.
By solving these challenges, PoT-quantized DNNs to be deployed end-to-end on specialized accelerators.
End-to-end deployment enables the evaluation of different PoT quantization methods in terms of accuracy, performance and energy efficiency on specialized accelerators.

In general, different PoT quantization methods introduce varying numbers of shifts and additions to replace a single multiplication in the design of a processing element (PE).
For example, the methods from Przewlocka-Rus et al.~\cite{przewlocka-rus_power--two_2022} and QKeras~\cite{qkeras} use a single shift operation to replace a multiplication.
Other methods, such as APoT~\cite{li_additive_2020} and MSQ~\cite{chang_mix_2021}, replace a multiplication operation by two shift operations and one addition.
Based on the quantization level, shift-values are different for both APoT and MSQ.
Additionally, previous research has also indicated that the inclusion of Zero-value as quantization level can increase the PE design complexity~\cite{przewlocka-rus_hwAccPoT_2023,li2023densshift}.
Such variations in PoT quantization methods introduce different complexities in hardware design, including those related to the PEs.
How this PE design complexity translates to performance, energy efficiency, and resource utilization when running DNNs on custom hardware accelerators remains to be explored.
Moreover, the current lack of integration between PoT-quantized models and inference frameworks, as well as custom hardware accelerators, makes it difficult to evaluate the benefits of PoT-quantized DNNs on resource-constrained edge devices.

To bridge this gap, we propose \texttt{PoTAcc}, a comprehensive pipeline for the end-to-end acceleration and evaluation of PoT-quantized DNNs on resource-constrained edge devices (introduced in our earlier work~\cite{saha2024accelerating}).
We chose TFLite for the inference framework because it is widely used for deploying DNNs on edge devices and supports custom hardware accelerators using its delegate system.
The TFLite delegate system enables DNN inference on heterogeneous hardware accelerator while sharing a common external memory interface.
We show how different types of pre-trained PoT-quantized DNNs can be prepared for TFLite to run efficiently on our proposed hardware accelerators.

The contributions of this paper include the following:

\begin{enumerate}
    \item Efficient shift-PE designs for three different PoT quantization methods: i) QKeras~\cite{qkeras}, ii) MSQ~\cite{chang_mix_2021}, and iii) APoT~\cite{li_additive_2020}.
    
    \item Implementation of hardware accelerators for these PoT quantization methods on two FPGA platforms(PYNQ-Z2, Kria) using the corresponding shift-PE designs.
   
    \item Integration of training (PyTorch~\cite{paszke2017pytorch}, TensorFlow~\cite{abadi2016}) and inference (TFLite~\cite{tflite}) frameworks within our proposed open-source \texttt{PoTAcc} pipeline to enable end-to-end deployment of PoT-quantized DNNs on resource-constrained edge devices. 
    
    \item Evaluation of accuracy, performance and energy efficiency across a wide range of DNNs, including four CNNs and two Transformer-based models.
    Our proposed accelerator paired with a CPU achieves up to $3.6\times$ speedup and a 78.0\% energy reduction when compared to CPU-only execution for different PoT-quantized DNNs while maintaining the accuracy.
\end{enumerate}

\begin{figure}[t]
 \centering
 \includegraphics[width=0.90\columnwidth]{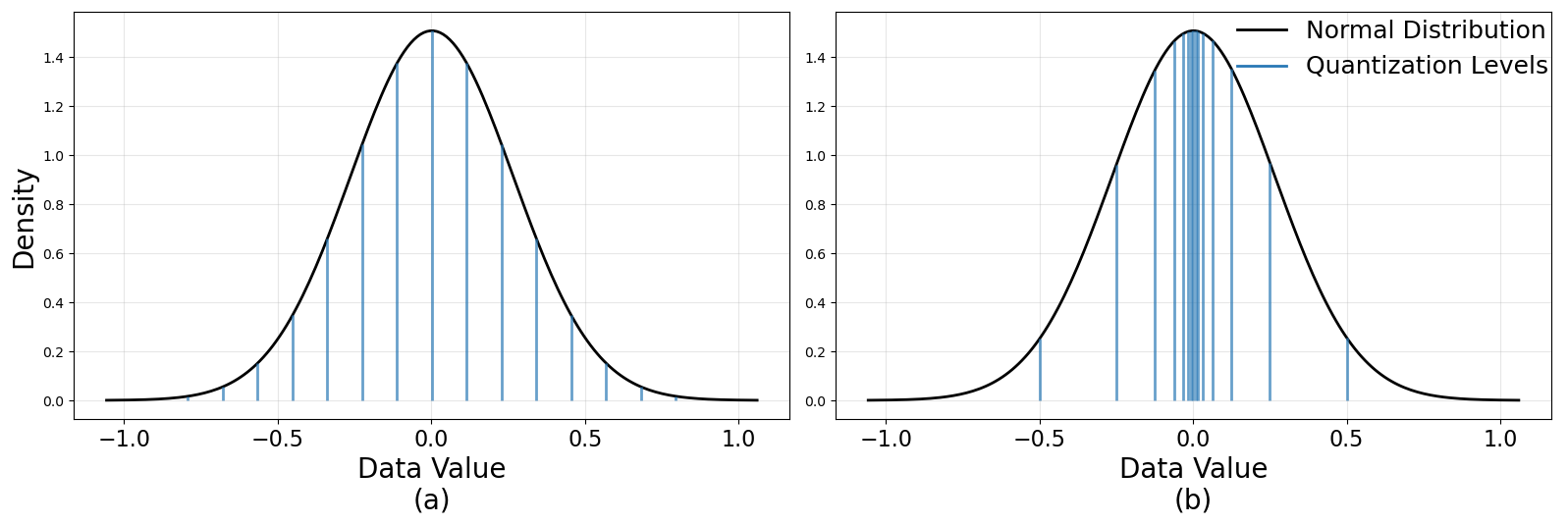}
 \caption{\label{fig:quantization_type} Quantization Types: (a) Unifrom Quantization; (b) Non-Uniform Quantization (PoT).}
\end{figure}

\section{Background and Motivation}
\label{sec:background and motivation}


\subsection{Power-of-Two Quantization}

Non-uniform quantization involves unequal distances between quantization levels (in uniform quantization, all distances are equal), and one form is PoT quantization~\cite{miyashita_convolutional_2016}, Figure~\ref{fig:quantization_type}.
The simplest PoT quantization method uses a single PoT term to express the quantization level to represent the original data(\!\!~\cite{miyashita_convolutional_2016,zhou_incremental_2017,elhoushi_deepshift_2021,przewlocka-rus_power--two_2022,jiang_jumping_2023,qkeras,li2023densshift}); more complex methods can use a combination of multiple PoT terms (\!\!~\cite{gudovskiy_shiftcnn_2017,vogel_bit-shift-based_2019,li_additive_2020,liu_optimize_2020,ansari_improved_2021,chang_mix_2021,gong_elastic_2021,xia_energy-and-area-efficient_2023}).
Single-PoT-term-based quantized weights, ${Q}_{W}$, can be represented as shown in Equation~\ref{eq:single-PoT-quantization}, where $b$ is the quantization bit-width, $\alpha$ is a scaling factor and \new{$pot\_float$ is the PoT quantization level in the floating-point format.}
\begin{equation}
\label{eq:single-PoT-quantization}
    {Q}_{W}=\alpha \times \new{pot\_float}
\end{equation} 
\begin{equation*}
    {Q}_{W}=\alpha \times\left\{0, \pm 2^{-2^{b-1}+1}, \pm 2^{-2^{b-1}+2}, \ldots, \pm 2^{-1}, \pm 1\right\}
\end{equation*} 
As an example of single-PoT-term-based PoT quantization, QKeras~\cite{qkeras} can be mentioned, a library from TensorFlow, provides various quantization methods including PoT quantization.
One of the characteristics of single-PoT-term-based quantization is that the density of the quantization levels increases around zero with the increase in quantization bit-width.
APoT~\cite{li_additive_2020} defined this as a \textit{rigid resolution}, arguing that instead of increasing the number of quantization levels near zero, it is better to have more quantization levels in the higher range.
APoT introduced multi-PoT-term-based quantization where the number of PoT terms per quantization level is determined by the bit-width.
For $b$ bit-width PoT quantization, they set $k=2$, where $k$ is the bit-width per PoT term.
For the signed version, $b^\prime = b-1$.
When $b^\prime = kn$, where $n$ is a positive integer, PoT terms are defined as shown in Equation~\ref{eq:apot-quantization-even}: 
\begin{equation}
\label{eq:apot-quantization-even}
Q_W =\alpha \times \new{pot\_float} = \alpha \times  \sum_{i=0}^{n-1} q_i ,
\end{equation}
\begin{equation*}
q_i \in 
\left\{
  0, \;
  \tfrac{1}{2^{i}}, \;
  \tfrac{1}{2^{\,i+n}}, \;
  \ldots, \;
  \tfrac{1}{2^{\,i+(2^{k}-2)n}}
\right\}.
\end{equation*} 
When $b^\prime = kn+1$, where $n$ is a positive integer, PoT terms are defined as shown in Equation~\ref{eq:apot-quantization-odd}: 
\begin{equation}
\label{eq:apot-quantization-odd}
    Q_W =\alpha \times \new{pot\_float} = \alpha \times \sum_{i=0}^{n-1} q_i + \tilde{q},
\end{equation}
\begin{equation*}
q_i \in 
\left\{ 
   0, \; \tfrac{1}{2^{i+1}}, \; \tfrac{1}{2^{\,i+1+n}}, \; \tfrac{1}{2^{\,i+1+2n+1}}
\right\}, 
\qquad
\tilde{q} \in 
\left\{ 
   0, \; \tfrac{1}{2^{\,2n+1}}
\right\}.
\end{equation*}
Later, MSQ~\cite{chang_mix_2021} proposed a double-PoT-term-based quantization where two PoT terms represent each quantization level, but the number of bits assigned to each PoT term depends on the total bit-width.
For $b$ bit-width PoT quantization, 1 bit is used for the sign.
MSQ~\cite{chang_mix_2021} can be represented by Equation~\ref{eq:msq-quantization}: 

\begin{equation}
\label{eq:msq-quantization}
    {Q}_{W}=\alpha \times \new{pot\_float} =\alpha \times\left\{\pm\left({q}_{0}+{q}_{1}\right)\right\}
\end{equation}
\begin{equation*}
\begin{array}{l}
q_{0} \in\left\{0, \frac{1}{2^{2^{b_{1}}-1}}, \frac{1}{2^{2^{b_{1}}-2}}, \ldots, \frac{1}{2}\right\}, \\
q_{1} \in\left\{0, \frac{1}{2^{2^{b_{2}}-1}}, \frac{1}{2^{2^{b_{2}}-2}}, \ldots, \frac{1}{2}\right\}
\end{array}
\begin{array}{l}
b^\prime=b_{1}+b_{2}, \\
b_{1} \geq b_{2}
\end{array}
\end{equation*} 
where ${q}_{0}$ and ${q}_{1}$ are PoT terms of each quantization level.

PoT terms enable the use of bit-shift operations to implement multiplications, since shifting by the log of a PoT-term, i.e., a shift-term, is equivalent to multiplying by the PoT-term.
For our proposed \texttt{PoTAcc} pipeline, we chose three different PoT quantization methods (QKeras~\cite{qkeras}, APoT~\cite{li_additive_2020}, and MSQ~\cite{chang_mix_2021}) to demonstrate the versatility of the pipeline.
We opted for using 4-bit PoT quantization for weights since it is a commonly used configuration for low-bit PoT quantization methods~\cite{przewlocka-rus_power--two_2022,li_additive_2020,chang_mix_2021}.
However, our pipeline is not limited to 4-bit PoT quantization, and it can be extended to support other bit-widths (e.g., 2/3-bit) as well by designing the corresponding shift-based PEs and accelerators.
Since our pipeline is built on TFLite, which supports only 8-bit uniform quantization for activations, we adopt this default quantization scheme for all activations.

\begin{table*}[t]
\centering
\caption{PoT values in $pot\_float$ and $pot\_int$ formats.}
\label{tab:potSchemesCombined}
\resizebox{0.75\textwidth}{!}{%
\begin{tabular}{|l|l|l|l|l|}
\hline
\multirow{2}{*}{\textbf{PoT Method}} 
& \multicolumn{2}{c|}{\textbf{\new{$pot\_float$}}} 
& \multicolumn{2}{c|}{\textbf{\new{$pot\_int$}}} \\ \cline{2-5}
& \textbf{1st PoT Term} & \textbf{2nd PoT Term} & \textbf{1st PoT Term} & \textbf{2nd PoT Term} \\ \hline
QKeras~\cite{qkeras}         
& \(\pm2^{-1}\) to \(\pm2^{-8}\) & NA 
& \(\pm2^{7}\) to \(\pm2^{0}\) & NA \\ \hline
MSQ~\cite{chang_mix_2021}    
& 0, \(\pm2^{-1}\), \(\pm2^{-2}\), \(\pm2^{-3}\) & 0, \(\pm2^{-1}\)   
& 0, \(\pm2^{2}\), \(\pm2^{1}\), \(\pm2^{0}\) & 0, \(\pm2^{2}\) \\ \hline
APoT~\cite{li_additive_2020} 
& 0, \(\pm2^{-1}\), \(\pm2^{-2}\), \(\pm2^{-4}\) & 0, \(\pm2^{-3}\)   
& 0, \(\pm2^{3}\), \(\pm2^{2}\), \(\pm2^{0}\) & 0, \(\pm2^{1}\) \\ \hline
\end{tabular}%
}
\end{table*}

For 4-bit PoT quantization, the PoT terms listed in Table~\ref{tab:potSchemesCombined} under \new{$pot\_float$} quantization level are calculated using Equations~\ref{eq:single-PoT-quantization},~\ref{eq:msq-quantization}, and~\ref{eq:apot-quantization-odd}.
Note that for QKeras, we remove $0$ as a quantization level to demonstrate the optimization in the corresponding shift-PE design that has been indicated by previous research~\cite{przewlocka-rus_hwAccPoT_2023,li2023densshift}.
During training, the quantization levels defined as \new{$pot\_float$} are employed, whereas for inference weights are transformed into their corresponding integer representations (\new{$pot\_int$, see Section~\ref{subsec:shift_pe_design}}) to facilitate shift-based operations.
This conversion enables integer multiplications to be efficiently replaced by \new{left} shift operations.


\subsection{Quantized Matrix multiplication (QMM)}

One of the goals of quantization is to reduce the computational cost of DNN models. To achieve this, we first need to clarify how matrix multiplication (MM) is performed in the quantized domain.
We are focusing on MM operations because the targeted DNNs consist of convolutional and fully-connected layers, which can be translated to MM.
The work of Jacob et al.~\cite{jacob_2018_CVPR} describes how to perform quantized matrix multiplication (QMM) using 8-bit uniform quantization (Equation~\ref{eq:qmmm-8bit}).

Let $O \in \mathbb{R}^{m \times n}$, $W \in \mathbb{R}^{n \times k}$, and $A \in \mathbb{R}^{m \times k}$ denote the output, weight, and activation matrices, respectively. 
Let $b \in \mathbb{R}^{n}$ denote the bias vector. 
In general, any real value $r \in \mathbb{R}$ can be quantized as $r = S (q - Z)$, where $q$ is the quantized integer value, $S$ is the quantization scale, and $Z$ is the zero-point. 
Using this scheme, the QMM with bias can be expressed as shown in Equation~\ref{eq:qmmm-8bit}: 
\begin{equation}
\label{eq:qmmm-8bit}
O = W A + b
\end{equation}
\begin{equation*}
    S_o(q_o - Z_o) = S_W S_A (q_W - Z_W)(q_A - Z_A) + S_b (q_b - Z_b)
\end{equation*} 
Using symmetric quantization for weights and bias (i.e., $Z_W = 0$ and $Z_b = 0$) and bias scale $S_b = S_W S_A$, the simplification of the equation leads to Equation~\ref{eq:qmmm-8bit-symmetric}: 
\begin{equation}
    \label{eq:qmmm-8bit-symmetric}
        q_o = \frac{S_W S_A}{S_o} \Big( q_W q_A  + (q_b - q_W Z_A) \Big) + Z_o
\end{equation} 
Here, $q_W q_A$ denotes the core matrix multiplication in the quantized domain. 
The term $(q_b - q_W Z_A)$ represents a precomputed offset that is added to each output element. 
Finally, the factor $\tfrac{S_W S_A}{S_o}$ rescales the result back into the quantized integer domain, with the output zero-point $Z_o$ applied at the end.
TFLite follows the same QMM approach as described in Equation~\ref{eq:qmmm-8bit-symmetric}.


\subsection{Motivation for \texttt{PoTAcc}}

General-purpose processors such as CPUs and GPUs typically lack native kernel support for the shift-based MM required by PoT quantization. 
Although DenseShift~\cite{li2023densshift} demonstrates the feasibility of executing PoT-weighted DNN models with FP16 activations and low-bit (2 or 3-bit) weights on CPUs by replacing multiplications with additions, such implementations do not generalize to models with quantized activations (e.g., 8-bit or 4-bit). 
Consequently, prior work~\cite{chang_mix_2021} has advocated to use dedicated FPGA-based accelerators for inference of PoT-quantized DNN models, particularly in edge computing scenarios. 

However, deploying PoT-quantized DNN models has remained challenging from an end-to-end perspective, primarily due to the absence of inference frameworks with native support for PoT quantization.
Therefore, evaluating a PoT-quantized DNN model on a custom accelerator from an end-to-end perspective, particularly in terms of latency and energy consumption, remains a significant challenge.
To the best of our knowledge, prior work~\cite{li_additive_2020,gudovskiy_shiftcnn_2017,xia_energy-and-area-efficient_2023,chang_mix_2021} has not found a comprehensive solution that simultaneously addresses model preparation, accelerator design, and deployment within an inference framework, thereby enabling end-to-end evaluation of PoT-quantized DNN models on custom hardware accelerators.

Every time a new quantization method emerges, it typically requires designing or modifying dedicated hardware to test its performance and energy efficiency, adding significant overhead to the research and development process.
Moreover, the lack of open-source accelerator designs further complicates this effort, as researchers cannot easily adopt, adapt, or build upon existing implementations to support new quantization strategies.

The goal with our \texttt{PoTAcc} pipeline is to create an integrated flow that connects a training framework, capable of training models with a specific PoT quantization method, with an inference framework that supports the design of accelerators tailored to those quantization methods.
Our \texttt{PoTAcc} pipeline supports model preparation steps to convert trained DNN models into a form compatible with the inference framework, ensuring a seamless transition from training to deployment.
Importantly, the accelerator designs in \texttt{PoTAcc} will be open-source, enabling other researchers to adopt, extend, or even design new accelerators from scratch based on their own PoT quantization strategies.
To the best of our knowledge, the \texttt{PoTAcc} pipeline is the first to prioritize both open-source availability and reusability in the design of shift-based accelerators for PoT-quantized DNN model inference.

\section{Hardware Analysis for \texttt{PoTAcc}}
\label{sec:analysis}

One of the goals of the \texttt{PoTAcc} pipeline is to generate efficient hardware accelerators for different PoT quantization methods.
In this section, we analyze the hardware challenges specific to PoT quantization methods.
We also focus on creating a generalized accelerator design that can be modified to support different PoT quantization methods across various types of DNNs (e.g., CNNs, Transformers, etc.).
At the same time, we focus on designing our accelerator to be scalable and support different FPGA platforms.


\begin{figure*}[t]
 \centering
 \includegraphics[width=0.95\textwidth]{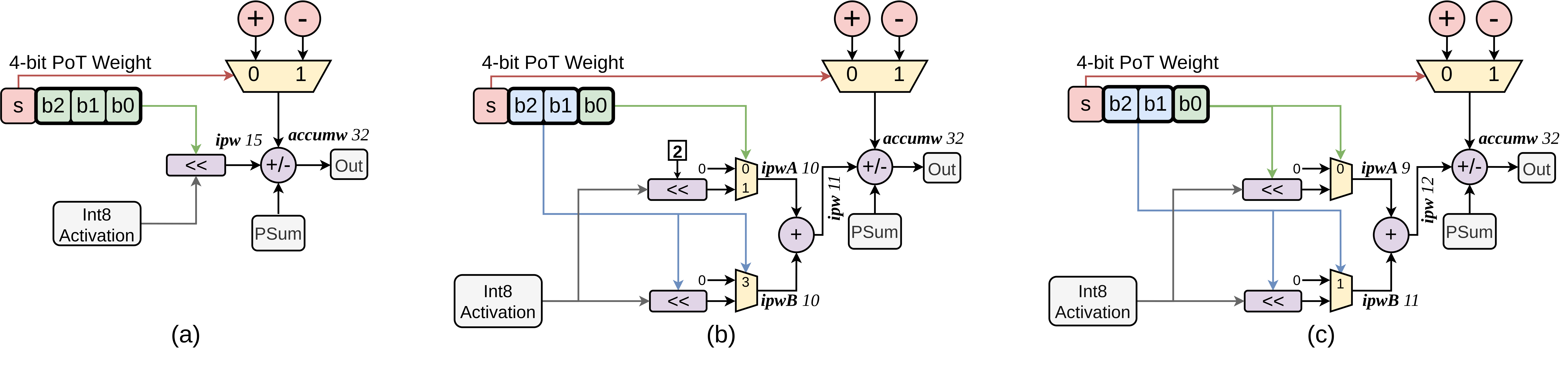}
 \caption{\label{fig:shift_pe_design} Shift-PE design: (a) 8\_4\_pot\_QKeras; (b) 8\_4\_pot\_MSQ; (c) 8\_4\_pot\_APoT.}
\end{figure*}

\subsection{Shift-PE Design}
\label{subsec:shift_pe_design}

We begin by analyzing the hardware cost of a \new{shift-PE}, which is fundamental to each PoT quantization method (QKeras~\cite{qkeras}, MSQ~\cite{chang_mix_2021}, and APoT~\cite{li_additive_2020}).
Table~\ref{tab:potSchemesCombined} shows the $pot\_float$ and $pot\_int$ representation of the PoT terms for each \new{4-bit} quantization method.
The easiest way to convert float-based PoT terms to integer-based PoT terms is to divide them by the smallest PoT value (except \(0\)) among the terms.
For example, QKeras~\cite{qkeras} has $pot\_float$ values from \(2^{-1}\) to \(2^{-8}\) including both PoT terms.
Dividing by the smallest $pot\_float$ value (i.e., \(2^{-8}\)) to all $pot\_float$ values, we get the $pot\_int$ values from \(2^7\) to \(2^0\).
A similar kind of conversion was also discussed in MSQ~\cite{chang_mix_2021}.
Since our activations use 8-bit integers, the \new{shift-PE} design depends on the $pot\_int$ representation of the PoT terms.

The shift-value can be represented as \(x\) in the \(2^x\) integer representation of the PoT term, where the PoT term is not zero.
\new{To design the shift-PE, we need to encode the shift values and the sign of the PoT terms in $pot\_int$ format within 4 bits.
In our design (see Figure~\ref{fig:shift_pe_design}), we reserve the most significant bit (MSB) for the sign bit.
For the QKeras~\cite{qkeras} method (see Figure~\ref{fig:shift_pe_design}(a)), we represent the shift values using 3 bits, which can represent values from \(0\) to \(7\) (i.e., PoT terms \(2^0\) to \(2^7\)).
If a PoT term is zero, we need to handle it as a special case, denoted as \(\eta\).
Table~\ref{tab:potSchemesCombined} shows that both the MSQ~\cite{chang_mix_2021} and APoT~\cite{li_additive_2020} methods contain zero values in their PoT terms in the $pot\_int$ representation.
The shift values and \(\eta\) are encoded in 4 bits for both methods, but the encoding scheme differs for each method (as shown in Figures~\ref{fig:shift_pe_design}(b) and (c)).
We reserve the 2 bits after the sign bit for the first PoT term, which can represent four possible values (\(0\) to \(3\)).
For MSQ, the shift values and \(\eta\) are encoded as follows: \(0\) to \(0\), \(1\) to \(1\), \(2\) to \(2\), and \(3\) to \(\eta\) (representing the special case).
For APoT, the encoding is as follows: \(0\) to \(0\), \(1\) to \(\eta\), \(2\) to \(2\), and \(3\) to \(3\).
For the second PoT term, we use the remaining least significant bit (LSB) to represent its shift value or \(\eta\), which can have two possible values.
For both MSQ and APoT, we map \(0\) to \(\eta\) and \(1\) to the corresponding shift value (\(2\) for MSQ and \(1\) for APoT).}

Based on the encoding scheme, shift-PE designs for the three PoT quantization methods are shown in Figure~\ref{fig:shift_pe_design}.
Two important things to notice here: (1) to handle the special case \(\eta\), extra hardware (e.g., a decoder mux) is required; (2) the bit-width of the intermediate product\new{, $ipw$,} depends on the maximum shift-value of the PoT quantization method, which can also affect the accumulator bit-width\new{, $accuw$.
The size of $accuw$ depends on the DNN layer size.
In our shift-PE designs, we keep $accuw$ at 32 bits for all quantization methods.
We use these shift-PE designs to generate accelerators for the corresponding PoT quantization methods.}

A similar design notation was used in previous works~\cite{przewlocka-rus_power--two_2022,li2023densshift}.
The key difference between QKeras~\cite{qkeras} and the method by Prezwlocka et al.~\cite{przewlocka-rus_power--two_2022} is that QKeras does not have \(0\) as a PoT term.
Therefore, the proposed shift-PE design for QKeras does not need a decoder mux to handle the special case \(\eta\).
One of the design choices made by the previous work~\cite{przewlocka-rus_power--two_2022} when comparing shift-PE designs between their PoT method and APoT, was to keep the intermediate product width ($ipw$) 12 bits and the accumulator width ($accuw$) 16 bits for both methods.
However, for single-shift-based PoT quantization methods such as QKeras~\cite{qkeras}, or the method by Przewlocka-Rus et al.~\cite{przewlocka-rus_power--two_2022}, $ipw$ is equal to the sum of the activation bit-width and maximum shift-value.
For an 8-bit activation and a maximum shift-value of 7 for QKeras (derived from $pot\_int$ values at Table~\ref{tab:potSchemesCombined})\new{,} $ipw$ is 15 bits.
\new{Note that} $ipw$ can be similarly reduced for APoT and MSQ.


\subsection{Vector-MAC (VMAC) Accelerator}

We aim to design an accelerator capable of supporting TFLite~\cite{tflite} models for inference. 
To this end, we employ the SECDA (SystemC Enabled Co-design of DNN Accelerators) methodology~\cite{haris_secda_2021} within the SECDA-TFLite~\cite{haris_secda_tflite_2023} toolkit for developing accelerators that target both $int8$ and PoT-quantized model inference.
Our work builds upon the previously proposed Vector MAC (VMAC) accelerator~\cite{haris_secda_tflite_2023}, which accelerates convolutional layers for $int8$-quantized models, while delegating the remaining layers to the CPU. 
The original VMAC accelerator, designed exclusively for $int8$ quantization, has been further optimized to enhance performance based on the observations, Section~\ref{oberservations}.

\subsubsection{General description} 

The VMAC accelerator~\cite{haris_secda_tflite_2023} in its original design consists of \textit{instruction handler}, \textit{data handler}, \textit{scheduler}, and four General Matrix Multiply (\textit{GEMM}) units, where each \textit{GEMM} unit has its own post-processing unit (\textit{PPU}), Figure~\ref{fig:dma_weight_copy}(a).
In terms of memory organization, the VMAC accelerator utilizes an AXI-Stream interface to fetch data from external DDR3 memory to the accelerator, and subsequently returns the computed results back to the DDR3 memory. 
The accelerator is equipped with a large global buffer (\textit{GACT}) for storing activation data.
In contrast, each \textit{GEMM} unit maintains its own local activation (\textit{LACT}) and weight (\textit{LWGT}) buffers. 
The micro-ISA defined under \textit{instruction handler} controls the execution of the accelerator, where the \textit{data handler} loads activations into the global buffer, whereas weights are distributed to the local buffers of the \textit{GEMM} units. 
During computation, the \textit{scheduler} determines which activation data from the global buffer is dispatched to each \textit{GEMM} unit. 
The scheduling strategy ensures that each \textit{GEMM} unit processes 16 outputs in parallel. 

\new{Within} VMAC~\cite{haris_secda_tflite_2023} each \textit{GEMM} unit comprises 64 multiply-PEs, with each PE capable of processing 8-bit weights and activations. 
Data is read from the local buffers, 64 weights and 64 activations, and is simultaneously supplied to the 64 multiply-PEs, thereby maximizing throughput. 
The intermediate products are accumulated in 32-bit accumulators and subsequently processed by the \textit{PPU}, which re-quantizes the results back to 8 bits before writing them to DDR3 memory. 
Since the \textit{PPU} directly streams results back to DDR3 memory, the accelerator avoids the need for dedicated local output buffers and effectively hides the latency to copy the output.

\begin{figure}[t]
    \centering
        \includegraphics[width=1\columnwidth]{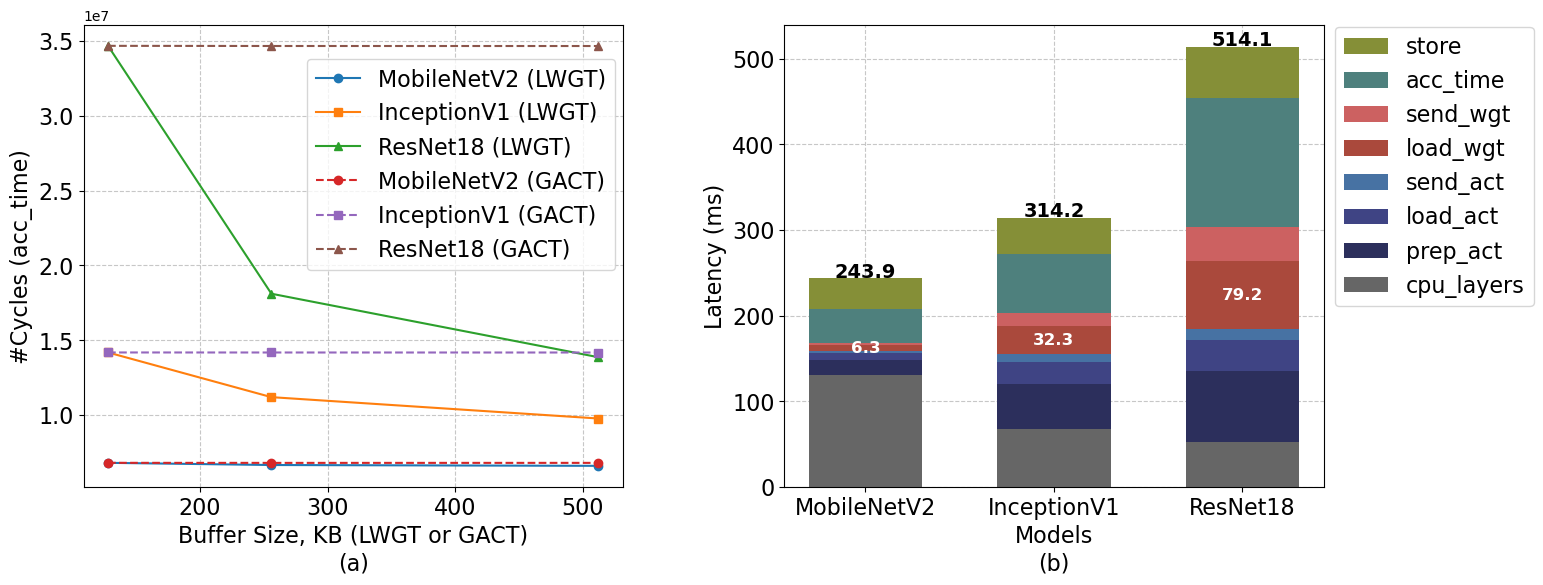}
        \caption{\label{fig:obsrvation-vmac-accel} VMAC Accelerator Observations: (a) effect of local memory capacity on simulated accelerator performance (acc\_time); (b) breakdown of inference time across different DNNs for VMAC~\cite{haris_secda_tflite_2023} on PYNQ-Z2 at 200MHz.}
\end{figure}

\subsubsection{Observations}
\label{oberservations}

The SECDA-TFLite toolkit~\cite{haris_secda_tflite_2023} enables end-to-end inference of $int8$-quantized models through the TFLite delegate system. 
Within this framework, the accelerator is registered as a delegate and whenever a targeted layer is encountered in the model graph during inference, the corresponding computation is offloaded to the accelerator. 
To improve the inference performance, it is necessary to understand how data flows through the delegate system and the behavior of the accelerator when the memory organization or capacity of the underlying hardware changes. 
Figure~\ref{fig:obsrvation-vmac-accel} presents our observations obtained using software-based and hardware-based profiling tools integrated within SECDA-TFLite.

Using software profiling tools in simulation, we varied the size of \textit{GACT} as well as \textit{LWGT} for three different DNNs.
While we varied one type of buffer size other buffer size kept at minimum value of 128 KB.
Figure~\ref{fig:obsrvation-vmac-accel}(a) shows that increasing the \textit{LWGT} reduces the number of clock cycles, whereas increasing the \textit{GACT} does not result in any significant improvement across different DNNs.
The reduction of number of clock cycles indicates the improvement of accelerator performance and it depends on the size of the DNNs.
For bigger DNN like ResNet18, for \textit{LWGT} at 512 KB, the number of clock cycles reduces more than 50\%.

Using hardware profiling tools, we obtained a detailed breakdown of inference performance for three representative DNNs executed on the PYNQ-Z2 FPGA board operating at 200~MHz, as shown in Figure~\ref{fig:obsrvation-vmac-accel}(b). 
The $prep\_{act}$ stage denotes the time required during inference to reorganize \new{activation} data from a single buffer into four buffers and perform the \texttt{im2col}~\cite{chellapilla2006high, jia2014learning} operation when necessary.
In the complete hardware design, four DMA blocks are employed to optimize data transfers between DDR3 memory and the accelerator. 
Initially, data resides in CPU memory, which is physically mapped to DDR3 memory and accessible only through the operating system. 
To offload data to the accelerator, data must first be copied from CPU memory into the DMA buffer; this transfer time is reported as $load^{*}$. 
Subsequently, data is transferred from the DMA buffer to the accelerator using four parallel DMAs, with the associated transfer time denoted as $send^{*}$. 
While activation data transfer must traverse the path CPU $\rightarrow$ DMA $\rightarrow$ accelerator, weight data can be preloaded into the DMA buffer prior to inference. 
Weight preloading avoids repeated $load\_{wgt}$ operations during execution, thereby reducing runtime overhead.
For example, for ResNet18, removing $load\_{wgt}$ can reduce overall latency by 15\%.
The $acc\_time$ metric represents the time consumed by the accelerator to process the input data and return the results to DDR3 memory.
\new{The variation in \textit{LWGT} size specifically impacts the $acc\_time$ metric. The simulated accelerator performance, measured in clock cycles and shown in Figure~\ref{fig:obsrvation-vmac-accel}(a), is proportional to $acc\_time$.}
Finally, the $store$ stage corresponds to the time required to copy output data from the DMA buffer back to CPU memory, including the overhead of performing data reorganization during this process.

\section{\texttt{PoTAcc} Pipeline}
\label{sec:pot_model_deployment}

\begin{figure*}[ht]
 \centering
 \includegraphics[width=0.75\textwidth]{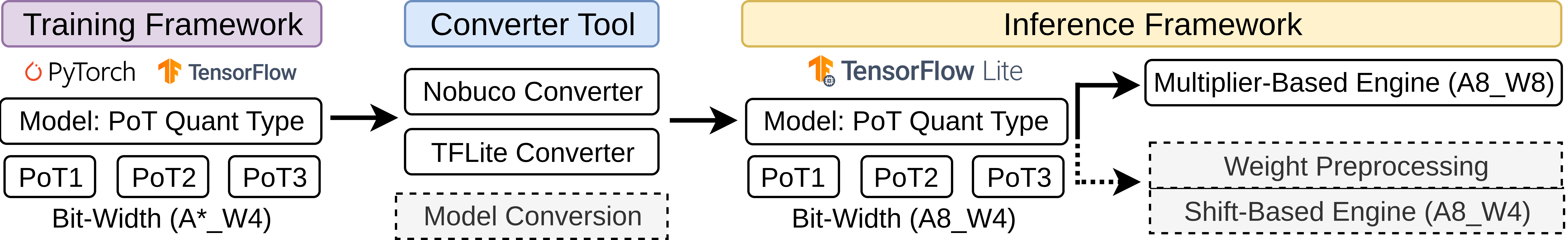}
 \caption{\label{fig:pot_deploy} \texttt{PoTAcc}: End-to-End Pipeline for PoT-quantized DNN Model Acceleration.
 }
\end{figure*}

\new{Figure~\ref{fig:pot_deploy} shows the main components of \texttt{PoTAcc}, our proposed open-source pipeline for end-to-end acceleration and evaluation of PoT-quantized DNNs on resource-constrained edge devices.}
Our contributions are marked by dotted lines and boxes.
The figure shows the integration of the training and inference frameworks for PoT-quantized DNNs.
\new{One of the novel aspects of our pipeline is that our integration process enables inference of trained PoT-quantized DNNs on custom hardware accelerator, while still maintaining the trained accuracy.}
In this integration process, we highlight our contributions in three key areas i) model conversion, ii) weight preprocessing, and iii) the design of shift-based accelerators; which are discussed in detail in the following subsections.


\subsection{Model Conversion}
\label{sec:model_conversion}

A PoT-quantized DNN model can be trained using standard deep learning frameworks such as PyTorch~\cite{paszke2017pytorch} or TensorFlow~\cite{abadi2016}.
Since the goal of \texttt{PoTAcc} is to enable inference through TFLite, any trained PoT-quantized model must be properly converted into the TFLite format.
\new{An incorrect model conversion can introduce various degrees of error~\cite{louloudakis_fetafix_2025}.
Our pipeline ensures that the conversion process preserves the accuracy of the trained PoT-quantized DNN models.}
To this end, we employ two model conversion tools: \texttt{Nobuco}~\cite{nobuco} and the native \texttt{TFLite Converter}~\cite{tflite}.
\texttt{Nobuco}~\cite{nobuco} translates PyTorch models into TensorFlow models, while the \texttt{TFLite Converter} subsequently transforms TensorFlow models into the TFLite representation.
The whole model conversion process is shown on first two sections of Figure~\ref{fig:pot_deploy}.

To illustrate this process, we consider two representative DNNs, ResNet20~\cite{He2016DeepRL} and MobileNetV2~\cite{mbv2}, trained on the CIFAR-10 dataset. 
ResNet20 \new{is trained with} 4-bit double-PoT-term-based APoT quantization, sourced from the open-source APoT repository implemented in PyTorch (Equation~\ref{eq:apot-quantization-odd}).
In contrast, MobileNetV2 is trained with QKeras library from TensorFlow and adopts 4-bit single-PoT-term quantization (Equation~\ref{eq:single-PoT-quantization}).

A PyTorch-based PoT-quantized model comprises a computation graph containing custom PoT quantization layers, along with trained dequantized weights and the corresponding quantization parameters stored in the state dictionary. 
However, the \texttt{Nobuco} converter does not support the direct translation of such custom quantization layers. 
To address this limitation, we modify the PyTorch model graph prior to conversion by replacing quantized convolutional and fully-connected (FC) layers with their standard counterparts. 
In addition, the dequantized weights stored in the state dictionary must be re-quantized using the forward function definition of the custom quantization layer and the associated quantization parameters. 
The resulting quantized weights can be expressed as $Q_{W}$, as defined in Equation~\ref{eq:apot-quantization-odd}. 
With the corrected model graph and quantized weights, the PyTorch model can then be successfully converted into a TensorFlow model using the \texttt{Nobuco} converter. 
Finally, the TensorFlow model is transformed into a TFLite model using the native \texttt{TFLite Converter}~\cite{tflite}.

For TensorFlow-based development, PoT-quantized models can be trained using the QKeras library~\cite{qkeras}, which provides native support for PoT quantization in DNNs. 
Specifically, QKeras supports single-shift-based PoT quantization as part of its quantization-aware training framework.
The quantized weights can be represented as $Q_{W}$, as defined in Equation~\ref{eq:single-PoT-quantization}. 
The conversion process for TensorFlow-based PoT-quantized models is straightforward; the native \texttt{TFLite Converter} directly transforms the TensorFlow model into the TFLite format, producing a fully optimized computational graph tailored for efficient $int8$ inference.

\begin{figure} [t]
 \centering
 \includegraphics[width=0.9\columnwidth]{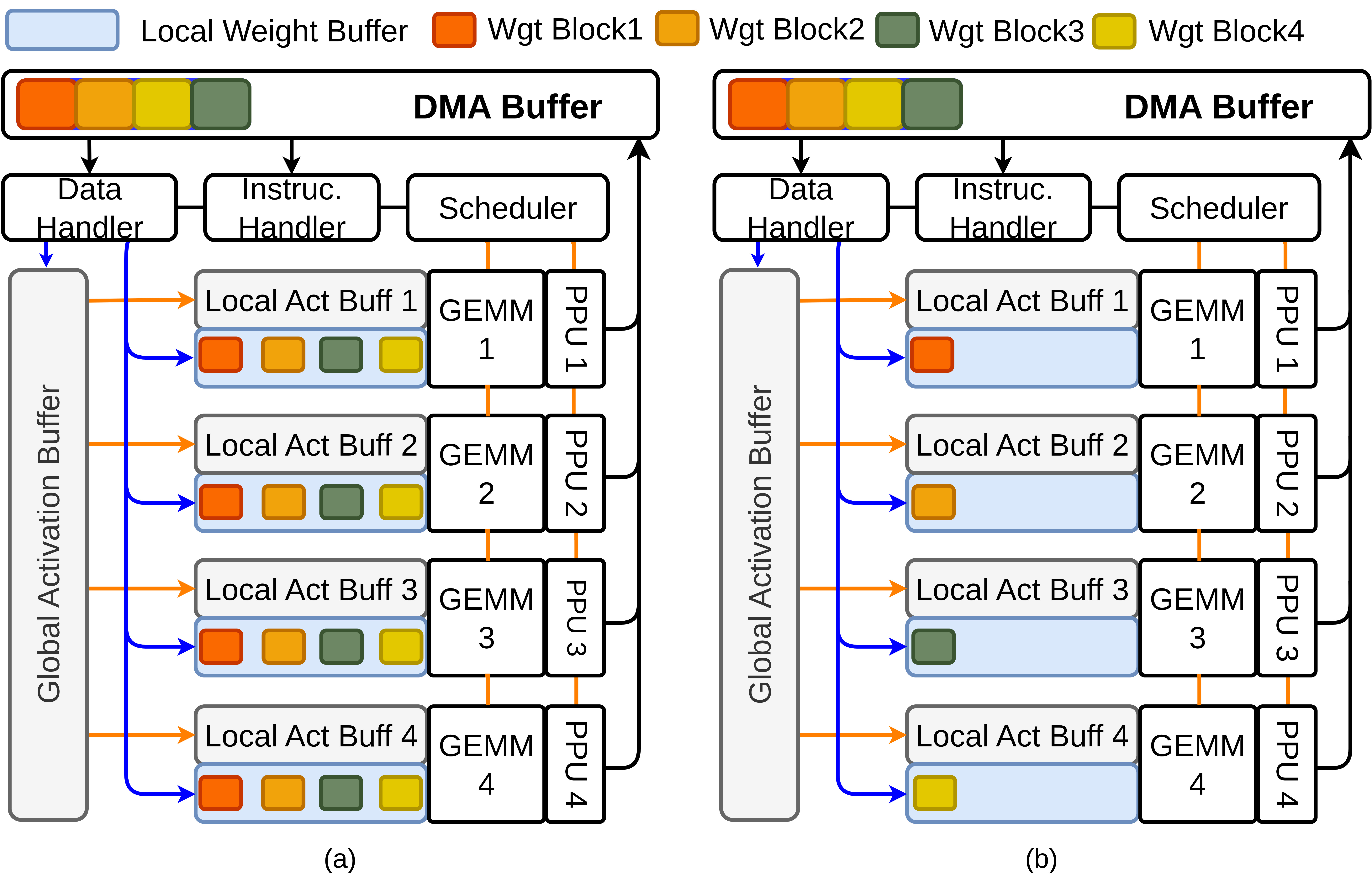}
  \caption{\label{fig:dma_weight_copy} Weight-copy strategy: (a) VMAC~\cite{haris_secda_tflite_2023}; (b) VMAC\_opt.}
\end{figure}


\subsection{Weight Preprocessing}
\label{sec:weight_preprocessing}

When a PoT-quantized DNN model is converted to TFLite format using the \texttt{TFLite converter}, the weights are quantized to $int8$ format, $q_{W}$, which is shown in Equation~\ref{eq:tflite-quantization}: 
\begin{equation}
  \label{eq:tflite-quantization}
q_{W} = \text{round}(\frac{Q_{W}}{S_{W}}) , S_{W} = \frac{max(|Q_{W}|)}{127}
\end{equation}
where \(Q_{W}\) are the PoT-quantized weights in $pot\_float$ multiplied by scale $\alpha$ and $S_{W}$ is the TFLite quantization scale factor.
To enable shift-based inference on the accelerator, we perform weight preprocessing to convert the $q_{W}$ weights to $pot\_int$ format, Table~\ref{tab:potSchemesCombined}.
Weight preprocessing has three steps 1) scale correction \new{2) encoding $pot\_int$ weights to 4-bit precision and 3) compressing the encoded values.}

Scale correction is used to bring the $q_{W}$ weights to the desired $pot\_int$ format.
For QKeras-based PoT quantization, the scale correction is not needed since the $q_{W}$ weights have the same range $\pm 127$ as the desired $pot\_int$ format.
However, for MSQ and APoT-based PoT quantization, the range in $pot\_int$ format is $\pm 10$ and $\pm 8$ respectively.
Therefore, scale correction is necessary for both methods.
Equation~\ref{eq:scale-correction} shows scale correction, where \(S_{pi}\) and \new{\(pot\_int\) are the scale and weights(i.e., PoT quantization level in integer format), respectively}: 
\begin{equation}
  \label{eq:scale-correction}
  Q_{W} \approx S_{W} q_{W} = (S_{W} C) \times \frac{q_{W}}{C} = S_{pi}\new{pot\_int} 
\end{equation}
\begin{equation*}
C = \frac{max(|q_{W}|)}{max(|\new{pot\_int}|)}
\end{equation*} 
The correction factor, \(C\), for each $q_W$ can be calculated beforehand since we know the quantized levels in three different formats, $pot\_float$, $int8$ and $pot\_int$ for each PoT quantization scheme.
Table~\ref{tab:weight_preprocessing_apot} lists the quantization levels in $pot\_float$, $int8$ and $pot\_int$ for the APoT PoT quantization scheme.
The quantization levels in $pot\_float$ and $pot\_int$ are derived from Table~\ref{tab:potSchemesCombined}, whereas the quantization levels in $int8$ are derived from Equation~\ref{eq:tflite-quantization}.
The correction of weight scales also requires the requantization of biases in QMM, Equation~\ref{eq:qmmm-8bit-symmetric}. 
Because, one of the assumptions in Equation~\ref{eq:qmmm-8bit-symmetric} are bias scale, \(S_b = S_W \times S_A\).
Therefore, when the weight scale is changed from \(S_W\) to \(S_{pi}\), the bias scale also needs to be changed to \(S_b = S_{pi} \times S_A\).

\begin{table*}[ht]
\centering
\caption{Weight preprocessing: mapping of floating-point and integer quantization levels for the \new{4-bit} APoT scheme.}
\label{tab:weight_preprocessing_apot}
\resizebox{0.95\textwidth}{!}{%
\begin{tabular}{|c|ccccccccccccccc|}
\hline
 \new{\textbf{Format}} & \multicolumn{15}{c|}{\textbf{Quantization Levels}} \\ \hline
\new{\textbf{$pot\_float$}}    & -0.625   & -0.5     & -0.375   & -0.25    & -0.1875   & -0.125   & -0.0625   & 0       & 0.0625  & 0.125   & 0.1875  & 0.25    & 0.375   & 0.5     & 0.625   \\ \hline
\new{\textbf{$int8$}}          & -127     & -102     & -76      & -51      & -38       & -25      & -13       & 0       & 13      & 25      & 38      & 51      & 76      & 102     & 127     \\ \hline
\new{\textbf{$pot\_int$}}      & -10      & -8       & -6       & -4       & -3        & -2       & -1        & 0       & 1       & 2       & 3       & 4       & 6       & 8       & 10      \\ \hline
\new{\textbf{$pot\_int^{e}$}}  & \new{-7} & \new{-6} & \new{-5} & \new{-4} & \new{-1}  & \new{-3} & \new{-0}  & \new{2} & \new{0} & \new{3} & \new{1} & \new{4} & \new{5} & \new{6} & \new{7} \\ \hline
\end{tabular}
}
\end{table*}

\new{Next, the sign and shift values of the quantized weights in $pot\_int$ format are encoded within 4 bits, as mentioned in Section~\ref{subsec:shift_pe_design}.
We refer to the encoded values as $pot\_int^{e}$, as shown in the last row of Table~\ref{tab:weight_preprocessing_apot} for the APoT scheme in sign magnitude format.
Finally, the encoded 4-bit $pot\_int^{e}$ weights are compressed to reduce the memory footprint. As each weight in the TFLite model is represented in an 8-bit format, the size reduction is achieved by packing two 4-bit weights into a single byte.}
The weight preprocessing step happens before the model inference starts.



\subsection{Accelerator Design and Optimizations}
\label{sec:accelerator_design_optimization}

We refer to the optimized VMAC~\cite{haris_secda_tflite_2023} accelerator as VMAC\_opt. 
The VMAC\_opt accelerator adopts the following optimizations to improve the resource utilization, performance and energy efficiency.

\subsubsection{Weight-copy Optimization}
\label{sec:weight_copy_optimization}

Based on the observations from Figure~\ref{fig:obsrvation-vmac-accel}(a), our objective is to increase the local weight buffer \textit{LWGT} capacity within each \textit{GEMM} unit. 
During the weight-copy process from the DMA to the accelerator, we observe that not all weight blocks in a tile are required by every \textit{GEMM} unit. 
Consequently, the weight-copy procedure can be optimized by transferring only the subset of weight blocks that are utilized by each \textit{GEMM} unit. 
This optimization is illustrated in Figure~\ref{fig:dma_weight_copy}(b). 
To enable this strategy, the scheduler code was modified to account for the selective weight-copy mechanism. 
With this optimized approach, the effective local weight buffer capacity of each \textit{GEMM} unit is increased by a factor of $4\times$.

\subsubsection{DMA Weight Preload Optimization}
\label{sec:dma_weight_preload_optimization}

Based on the second observation from Figure~\ref{fig:obsrvation-vmac-accel}(b), we introduce a DMA weight preloading optimization.
Within the software-side code of the delegate system, model weights are partitioned into four buffers and preloaded into the DMA memory during the \texttt{prepare()} function. 
This function is invoked once prior to inference to ensure that all weights are available in the DMA buffer throughout execution. 
It is therefore necessary to provision sufficient DMA buffer capacity to accommodate the complete set of weights for a given DNN model.
Note that weight preprocessing, as described in Section~\ref{sec:weight_preprocessing}, is also performed within the \texttt{prepare()} function.

\subsubsection{Support for Fully Connected (FC) Layers}

The original VMAC accelerator provided support exclusively for convolutional(CONV) layers. 
In the TFLite delegate implementation, we extended its functionality to also support FC layers. 
The key distinction between convolutional and FC layers arises from TFLite's aproach to quantization. 
While convolutional layers typically employ per-filter quantization, FC layers often rely on per-layer quantization. 
In per-filter quantization, a distinct scale is assigned to each filter, whereas in per-layer quantization, a single scale is applied across the entire layer. 
To enable support for FC layers, the \texttt{prepare()} function of the delegate is modified to replace the single scale with a vector of scales, where each element corresponds to a filter and is initialized by duplicating the original per-layer scale. 
Furthermore, unlike convolutional layers, FC layers do not require to perform \texttt{im2col}~\cite{chellapilla2006high, jia2014learning} operations during the $prep\_{act}$ stage.

\subsubsection{Scalability and Other Optimizations}

The original VMAC accelerator was designed without support for scaling up or down based on the available FPGA resources. 
In our SystemC-based implementation, we introduced a sub-module based \textit{GEMM} instantiation strategy, which enables the accelerator to be configured with 2, 4, 8, or 16 \textit{GEMM} units, thereby supporting resource scaling across different target boards.
In our implementation, we targeted two FPGA boards, the PYNQ-Z2~\cite{PYNQz2} and the Xilinx Kria~\cite{KRIA} platforms, and configured the accelerator with 4 and 8 \textit{GEMM} units, respectively.
The accelerator employs on-chip Block RAM (BRAM) to implement activation and weight buffers, with BRAM supporting a maximum data width of 36 bits. 
In the original VMAC design, both activation and weight buffers were implemented with a 32-bit data width. 
For FPGAs with larger resource capacity, such as the Xilinx Kria~\cite{KRIA} platform, Ultra RAM (URAM) is available, providing support for data widths up to 72 bits. 
To ensure portability across FPGA architectures, we modified the implementation of the local weight and activation buffers to operate with a 64-bit data width, which is compatible with both BRAM- and URAM-based memory systems.
Additionally, we simplified the $quantizer\_func()$ within the \textit{PPU} by retaining only a 64-bit multiplier and adder, along with a 32-bit shifter.

\subsubsection{Shift-based Accelerator}

To enable inference of PoT-quantized DNN models, we replaced the multiplication-based processing elements (mult-PEs) in the VMAC\_opt accelerator with shift-based processing elements (shift-PEs), resulting in the Vector-Shift Accumulator (VSAC) accelerator. 
The designs of the shift-PEs corresponding to different PoT quantization methods are illustrated in Figure~\ref{fig:shift_pe_design}. 
In the baseline VMAC design, the intermediate product width of a mult-PE is 16 bits. In contrast, in the shift-PEs the intermediate product width varies depending on the PoT quantization scheme. 
Shift operations are performed on $int8$ activations, which are represented in two’s complement format. 
Instead of applying sign correction during each shift operation, the sign of the intermediate product is handled during the accumulation phase. Depending on the weight sign bit, the intermediate product is either added to or subtracted from the accumulator. 

Another difference arises from the weight precision; mult-PEs utilize 8-bit weights, whereas shift-PEs operate on 4-bit weights. 
The reduced precision increases the effective capacity of the local weight buffers and decreases the number of weight tiles that must be transferred from the DMA to the accelerator. 
In the FPGA implementation, shift operations are mapped to LUTs, whereas multiplications are mapped to either LUTs or DSP blocks. 
To ensure a fair comparison between the VMAC\_opt and the VSAC accelerator, mult-PEs were also mapped to LUTs whenever sufficient LUT resources were available for the selected number of \textit{GEMM} units.

\section{Evaluation}
\label{sec:evaluation}


\begin{table*}[hbt]
\centering
\caption{Resource utilization (\%) comparison of full hardware design including DMAs and accelerator.}
\label{tab:resource_utilization}
\resizebox{0.8\textwidth}{!}{
\begin{tabular}{|l|c|c|c|c|c|c|c|c|c|}
\hline
\textbf{Acc Name} & \textbf{VMAC~\cite{haris_secda_tflite_2023}} & \multicolumn{2}{c|}{\textbf{VMAC\_opt}} & \multicolumn{2}{c|}{\textbf{VSAC\_QKeras}} & \multicolumn{2}{c|}{\textbf{VSAC\_MSQ}} & \multicolumn{2}{c|}{\textbf{VSAC\_APoT}} \\ \hline
\textbf{Quantization} & \multicolumn{3}{c|}{$int8$ (A8W8)}    & \multicolumn{6}{c|}{\new{$pot\_int^{e}$} (A8W4)}          \\ \hline
\textbf{Device}     & PYNQ-Z2 & PYNQ-Z2 & Kria  & PYNQ-Z2 & Kria   & PYNQ-Z2 & Kria   & PYNQ-Z2 & Kria \\ \hline
\textbf{Freq (MHz)} & 200     & 200     & 250   & 200     & 250    & 200     & 250    & 200     & 250  \\ \hline
\textbf{LUT}        & 94.5    & 89.2    & 83.5  & 67.8    & 61.2   & 69.9    & 63.4   & 73.3    & 65.1 \\ \hline
\textbf{FF}         & 57.5    & 43.4    & 29.8  & 42.5    & 28.0   & 46.1    & 32.2   & 46.7    & 32.3 \\ \hline
\textbf{DSP}        & 85.5    & 21.8    & 6.1   & 14.6    & 4.8    & 14.6    & 4.8    & 14.6    & 4.8  \\ \hline
\textbf{BRAM}       & 78.9    & 91.4    & 68.1  & 90.0    & 68.1   & 90.0    & 68.1   & 90.0    & 68.1 \\ \hline
\textbf{URAM}       & -       & -       & 100.0 & -       & 100.0  & -       & 100.0  & -       & 100.0  \\ \hline
\end{tabular}
}
\end{table*}

\begin{figure}[t]
    \centering
\includegraphics[height=0.15\textheight]{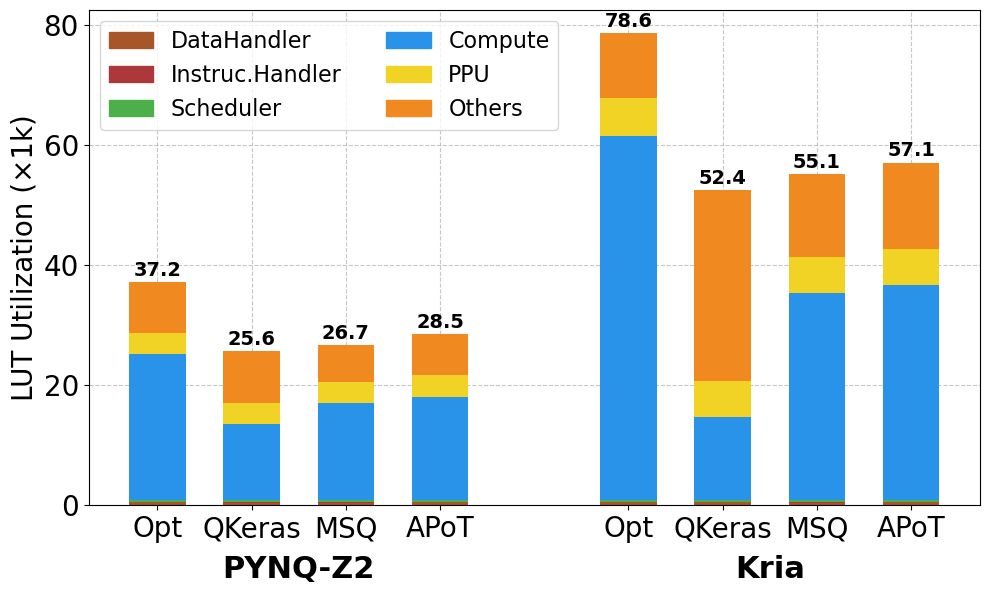}
    \caption{LUT utilization breakdown for different components of VMAC\_opt and VSAC accelerators on PYNQ-Z2 and Kria.}
    \label{fig:resource-breakdown}
\end{figure}

\subsection{Experimental Setup}

\subsubsection{Hardware Platforms}

Since our focus is on edge inference, we used the PYNQ-Z2~\cite{PYNQz2} and Kria KV260~\cite{KRIA} boards. 
Both boards feature an ARM CPU and an edge FPGA within the same System-on-Chip (SoC). 
The PYNQ-Z2 is equipped with a dual-core ARM Cortex-A9 CPU, while the Kria features a quad-core ARM Cortex-A53 CPU.
Their main difference is in the number of LUTs, DSPs, and FFs.
Both FPGAs have 630 Kilobytes (KB) of block-RAMs (BRAMs) but the Kria has an additional memory, 2.25 Megabytes (MB) of UltraRAM (URAM).
As discussed in Section~\ref{sec:accelerator_design_optimization}, we designed our shift accelerator, VSAC, using the SECDA methodology~\cite{haris_secda_2021}.
The SECDA-TFLite toolkit~\cite{haris_secda_tflite_2023} helped us to integrate the shift accelerator in TFLite to run and evaluate PoT-quantized DNNs within our \texttt{PoTAcc} pipeline.
\new{We used an NVIDIA RTX A6000 GPU with 48GB memory for training our PoT-quantized DNN models.}

\subsubsection{DNN Models and Dataset}

We selected a diverse set of DNNs, including convolutional neural networks (CNNs) and vision transformers (ViTs), to evaluate our \texttt{PoTAcc} pipeline.
\new{For} the CNNs, we chose MobileNetV2~\cite{mbv2}, InceptionV1~\cite{szegedy2015going}, ResNet18, ResNet20~\cite{He2016DeepRL}, and EfficientNet-Lite~\cite{tan2019efficientnet}.
For the ViTs, we selected ViT\_tiny and ViT\_small~\cite{dosovitskiy2020image}, as well as DeiT\_tiny and DeiT\_small~\cite{touvron2021training}.
We used ImageNet~\cite{ILSVRC15} and CIFAR-10~\cite{krizhevsky2009learning} datasets.

\subsubsection{DNN preparation}
\label{sec:dnn_prep}

\new{Two ImageNet-based DNNs (MobileNetV2 and ResNet18) and two CIFAR-10-based DNNs (MobileNetV2 and ResNet20) were trained using different PoT quantization methods, whereas the remaining DNNs were synthetically generated for the ImageNet dataset using all three PoT quantization methods (QKeras, MSQ, APoT).
During training, we kept activations at 32-bit precision, while weights were quantized using the 4-bit PoT quantization method in the $pot\_float$ representation, as described in Table~\ref{tab:potSchemesCombined}.
The weights of the first and last layers of the DNNs are quantized using the 8-bit uniform quantization method~\cite{jacob_2018_CVPR}, as commonly done in the literature~\cite{li_additive_2020, chang_mix_2021}.
All quantized weights are kept in 32-bit precision during training, and quantization is applied on-the-fly during the forward pass.
Training is performed for 20 epochs with an initial learning rate of 0.001, which is reduced by a factor of 10 after 5 and 15 epochs.
A Stochastic Gradient Descent (SGD) optimizer, with a momentum of 0.9 and a weight decay of $1e^{-4}$, is used for retraining, along with a batch size of 512.
The weights of the pre-trained floating-point models are used to initialize the training of the PoT-quantized DNN models.
After training, using model conversion (see Section~\ref{sec:model_conversion}), we convert the models to TFLite  format, (i.e., $int8$-quantized, 8-bit activations and 8-bit weights).
Later, we apply weight preprocessing (see Section~\ref{sec:weight_preprocessing}) to convert the weights of the PoT-quantized layers to the corresponding PoT representation ($pot\_int^{e}$), i.e., 8-bit activations and 4-bit weights.
Note that $int8$-quantized models are used to run on the CPU, VMAC~\cite{haris_secda_tflite_2023}, and VMAC\_opt accelerators, while weights in the $pot\_int^{e}$ representation are used to run on the corresponding VSAC accelerator.
Model execution on the accelerator occurs in a heterogeneous fashion, with the accelerator handling the PoT-quantized layers while the CPU processes the remaining layers.
} 

\new{Due to the lack of publicly available pre-trained PoT-quantized DNN models for the ImageNet dataset across different PoT quantization methods, we created synthetic PoT-quantized DNN models for our latency and energy evaluation.
The weights of the synthetic DNN models were quantized according to the $pot\_float$ representation described in Table~\ref{tab:potSchemesCombined} within the PyTroch training framework.}
\new{Subsequently, our models were converted into TFLite format using the \texttt{PoTAcc} pipeline.}
While our synthetic DNN models do not achieve state-of-the-art accuracy, they help us consistently evaluate the performance of our accelerators across different DNNs.

\subsubsection{Energy Measurement Setup}

For the PYNQ-Z2 platform, we used a USB-based power meter (AVHzY CT-3~\cite{avhzyct3}) to measure the power consumption of the entire board during inference.
For the Kria, we used the on-board power sensors to measure the power consumption.
Energy consumption, reported in joules/image, are calculated using Equation~\ref{eq:energy}.
$Power_{inference}$ is the average power consumption during inference, while $Power_{idle}$ is the average power consumption when the board is idle. 
\begin{equation}
    \text{Energy (J/image)} = \frac{Power (W) \times \text{Time (s)}}{\text{Number of images}}
    \label{eq:energy}
\end{equation}
\begin{equation*}
   Power (W) = Power_{inference} - Power_{idle}
\end{equation*}


\begin{table*}[t]
\centering
\caption{\new{Top-1(\%) Accuracy comparison across different stages of \texttt{PoTAcc}.}}
\label{tab:accuracy_test}
{\color{black}
\resizebox{\textwidth}{!}{%
\begin{tabular}{|c|c|c|c|c|c|c|c|}
\hline
\textbf{Dataset} & \textbf{Model(Quantization)} & \textbf{Framework} & \textbf{Precision(Format)} & \textbf{Stages} & \textbf{Hardware} & \textbf{Top-1(\%)} & \textbf{Top-1$\Delta$(\%)} \\
\hline
\multirow{6}{*}{CIFAR-10}
& \multirow{3}{*}{ResNet20 (APoT)}      & PyTorch                 & A32W32 ($pot\_float$) & Training             & GPU              & 92.6 & --  \\ \cline{3-8}
&                                       & \multirow{2}{*}{TFLite} & A8W8 ($int8$)         & Model Conversion     & CPU+VMAC\_opt    & 92.4 & -0.2 \\ \cline{4-8}
&                                       &                         & A8W4 ($pot\_int^{e}$)     & Weight Preprocessing & CPU+VSAC\_APoT   & 92.6 & 0.0 \\ \cline{2-8}
& \multirow{3}{*}{MobileNetV2 (QKeras)} & TensorFlow              & A32W32 ($pot\_float$) & Training             & GPU              & 93.9 & --  \\ \cline{3-8}
&                                       & \multirow{2}{*}{TFLite} & A8W8 ($int8$)         & Model Conversion     & CPU+VMAC\_opt    & 93.9 & 0.0 \\ \cline{4-8}
&                                       &                         & A8W4 ($pot\_int^{e}$)     & Weight Preprocessing & CPU+VSAC\_QKeras & 93.5 & -0.4 \\
\hline
\hline
\multirow{6}{*}{ImageNet}
& \multirow{3}{*}{ResNet18 (APoT)}      & PyTorch                 & A32W32 ($pot\_float$) & Training             & GPU              & 70.2 & --  \\ \cline{3-8}
&                                       & \multirow{2}{*}{TFLite} & A8W8 ($int8$)         & Model Conversion     & CPU+VMAC\_opt    & 69.2 & -1.0 \\ \cline{4-8}
&                                       &                         & A8W4 ($pot\_int^{e}$)     & Weight-Preprocessing & CPU+VSAC\_APoT   & 69.2 & -1.0 \\ \cline{2-8}
& \multirow{3}{*}{MobileNetV2 (APoT)}   & PyTorch                 & A32W32 ($pot\_float$) & Training             & GPU              & 71.3 & --  \\ \cline{3-8}
&                                       & \multirow{2}{*}{TFLite} & A8W8 ($int8$)         & Model Conversion     & CPU+VMAC\_opt    & 69.5 & -1.8 \\ \cline{4-8}
&                                       &                         & A8W4 ($pot\_int^{e}$)     & Weight-Preprocessing & CPU+VSAC\_APoT   & 69.4 & -1.9 \\
\hline
\end{tabular}%
}
}
\vspace{2pt}
\begin{flushleft}
\footnotesize{\new{Top-1$\Delta$(\%): Absolute Top-1 accuracy drop (in percentage points) with respect to the corresponding \textit{Training} stage.}}
\end{flushleft}
\end{table*}

\begin{table*}[t]
\centering
\caption{End-to-end evaluation: inference time (ms, speedup) and energy consumption (J/images, reduction) on PYNQ-Z2 and Kria platforms.
}
\label{tab:end-to-end-evaluation}
\resizebox{1.0\textwidth}{!}{
\begin{tabular}{|c|l|cccccc|cccccc|}
\hline
\multirow{2}{*}{\textbf{DNN}} & \multirow{2}{*}{\textbf{Hardware}} & 
\multicolumn{6}{c|}{\textbf{PYNQ-Z2 (CPU-2T)}} & 
\multicolumn{6}{c|}{\textbf{Kria (CPU-4T)}} \\ \cline{3-14}
 & & \textbf{$T_{conv}+T_{fc}$} & \textbf{$T_{other}$} & \multicolumn{2}{c}{\textbf{$Total Time$}} & \multicolumn{2}{c|}{\textbf{$Energy$}} 
   & \textbf{$T_{conv}+T_{fc}$} & \textbf{$T_{other}$} & \multicolumn{2}{c}{\textbf{$Total Time$}} & \multicolumn{2}{c|}{\textbf{$Energy$}} \\ \hline
 
\parbox[t]{6mm}{\multirow{6}{*}{\rotatebox[origin=c]{90}{\shortstack{MobileNetV2 \\ \cite{mbv2}}}}}
    & CPU                                & 225.7          & 130.3 & 356.0 & 1.0$\times$    & 0.90          & 00.0\% & \textbf{32.9}  & 32.7  & 65.6  & 1.0$\times$ & \textbf{0.04} & 00.0\% \\ 
    & CPU + VMAC~\cite{haris_secda_tflite_2023} & 115.6          & 130.0 & 245.6 & 1.4$\times$    & 0.61          & 32.0\% & --             & --    & --    & --          & --            & --    \\  
    & CPU + VMAC\_opt                    & 93.7           & 130.3 & 224.1 & 1.6$\times$    & 0.58          & 36.0\% & 77.5           & 32.7  & 110.2 & 0.6$\times$ & 0.06          & -52.2\% \\ 
    & CPU + VSAC\_QKeras                 & \textbf{91.8}  & 130.2 & 222.1 & 1.6$\times$    & \textbf{0.54} & 40.0\% & 77.1           & 33.3  & 110.4 & 0.6$\times$ & 0.05          & -24.2\% \\ 
    & CPU + VSAC\_MSQ                    & \textbf{92.4}  & 129.2 & 221.7 & 1.6$\times$    & \textbf{0.54} & 40.0\% & 77.2           & 34.7  & 111.9 & 0.6$\times$ & 0.05          & -25.1\% \\ 
    & CPU + VSAC\_APoT                   & \textbf{92.2}  & 129.7 & 221.9 & 1.6$\times$    & \textbf{0.54} & 40.0\% & 77.1           & 32.8  & 109.8 & 0.6$\times$ & 0.05          & -26.5\% \\ \hline

\parbox[t]{6mm}{\multirow{6}{*}{\rotatebox[origin=c]{90}{\shortstack{InceptionV1 \\ \cite{szegedy2015going}}}}} 
    & CPU                                & 718.1          & 67.2 & 785.3 & 1.0$\times$  & 2.65          & 00.0\% & 117.7          & 20.1  & 137.8 & 1.0$\times$ & 0.11          & 00.0\% \\
    & CPU + VMAC~\cite{haris_secda_tflite_2023} & 246.9          & 65.9 & 312.8 & 2.5$\times$  & 0.90          & 66.2\% & --             & --    & --    & --          & --            & --     \\ 
    & CPU + VMAC\_opt                    & 171.2          & 67.2 & 238.3 & 3.3$\times$  & 0.68          & 74.2\% & 91.6           & 20.1  & 111.7 & 1.2$\times$ & 0.09          & 20.6\% \\
    & CPU + VSAC\_QKeras                 & \textbf{162.2} & 67.3 & 229.5 & 3.4$\times$  & \textbf{0.63} & 76.3\% & \textbf{90.2}  & 20.0  & 110.2 & 1.3$\times$ & \textbf{0.06} & 45.0\% \\
    & CPU + VSAC\_MSQ                    & \textbf{163.6} & 67.2 & 230.8 & 3.4$\times$  & \textbf{0.64} & 76.0\% & \textbf{90.3}  & 20.0  & 110.3 & 1.2$\times$ & \textbf{0.06} & 45.0\% \\
    & CPU + VSAC\_APoT                   & \textbf{162.7} & 67.2 & 229.9 & 3.4$\times$  & \textbf{0.65} & 75.6\% & \textbf{90.4}  & 20.0  & 110.3 & 1.2$\times$ & \textbf{0.06} & 43.0\% \\ \hline

\parbox[t]{6mm}{\multirow{6}{*}{\rotatebox[origin=c]{90}{\shortstack{ResNet18 \\ \cite{He2016DeepRL}}}}}
    & CPU                                & 945.1          & 51.5 & 996.6 & 1.0$\times$     & 3.34          & 00.0\% & 146.2          & 19.6   & 165.8 & 1.0$\times$ & 0.16          & 00.0\% \\ 
    & CPU + VMAC~\cite{haris_secda_tflite_2023} & 462.9          & 51.6 & 514.5 & 1.9$\times$     & 1.41          & 57.7\% & --             & --     & --    & --          & --            & --     \\  
    & CPU + VMAC\_opt                    & 251.8          & 51.5 & 303.3 & 3.3$\times$     & 0.83          & 75.2\% & 109.1          & 19.6   & 128.7 & 1.3$\times$ & 0.11          & 33.0\% \\ 
    & CPU + VSAC\_QKeras                 & \textbf{220.3} & 52.9 & 273.2 & 3.6$\times$     & \textbf{0.74} & 78.0\% & \textbf{107.4} & 19.5   & 126.9 & 1.3$\times$ & \textbf{0.08} & 53.3\% \\ 
    & CPU + VSAC\_MSQ                    & \textbf{220.8} & 52.9 & 273.7 & 3.6$\times$     & \textbf{0.75} & 77.6\% & \textbf{107.7} & 19.6   & 127.3 & 1.3$\times$ & \textbf{0.08} & 52.8\% \\ 
    & CPU + VSAC\_APoT                   & \textbf{220.7} & 52.9 & 273.6 & 3.6$\times$     & \textbf{0.75} & 77.6\% & \textbf{107.8} & 19.6   & 127.3 & 1.3$\times$ & \textbf{0.08} & 51.4\% \\ \hline

\parbox[t]{6mm}{\multirow{6}{*}{\rotatebox[origin=c]{90}{\shortstack{EfficientNet-L \\ \cite{tan2019efficientnet}}}}}
    & CPU                                & 1405.3          & 653.3 & 2058.6 & 1.0$\times$  & 5.08          & 00.0\% & \textbf{209.4} & 88.8 & 298.2 & 1.0$\times$ & 0.26          & 00.0\% \\ 
    & CPU + VMAC~\cite{haris_secda_tflite_2023} & 538.0           & 655.4 & 1193.4 & 1.7$\times$  & 2.88          & 43.3\% & --             & --   & --    & --          & --            & --     \\  
    & CPU + VMAC\_opt                    & 395.5           & 653.3 & 1048.9 & 2.0$\times$  & 2.63          & 48.2\% & 315.4          & 88.8 & 404.1 & 0.7$\times$ & 0.27          & -2.2\% \\ 
    & CPU + VSAC\_QKeras                 & \textbf{378.7}  & 637.0 & 1015.7 & 2.0$\times$  & \textbf{2.63} & 48.2\% & 313.1          & 88.9 & 402.1 & 0.7$\times$ & \textbf{0.20} & 22.4\% \\ 
    & CPU + VSAC\_MSQ                    & \textbf{381.8}  & 654.7 & 1036.5 & 2.0$\times$  & \textbf{2.52} & 50.4\% & 313.0          & 89.0 & 401.9 & 0.7$\times$ & \textbf{0.21} & 21.1\% \\ 
    & CPU + VSAC\_APoT                   & \textbf{381.5}  & 657.0 & 1038.5 & 2.0$\times$  & \textbf{2.52} & 50.4\% & 312.9          & 87.6 & 400.5 & 0.7$\times$ & \textbf{0.21} & 18.3\% \\ \hline

\parbox[t]{6mm}{\multirow{6}{*}{\rotatebox[origin=c]{90}{\shortstack{DeiT\_tiny \\ \cite{touvron2021training}}}}}
    & CPU                                & 520.8          & 1007.1 & 1527.9 & 1.0$\times$  & 3.56          & 00.0\% & 76.3          & 332.3 & 408.6 & 1.0$\times$ & 0.23          & 00.0\% \\ 
    & CPU + VMAC~\cite{haris_secda_tflite_2023} & --             & --     & --     & --           & --            & --     & --            & --    & --    & --          & --            & --     \\  
    & CPU + VMAC\_opt                    & 93.1           & 1007.1 & 1100.2 & 1.4$\times$  & 2.56          & 28.3\% & 69.5          & 332.3 & 401.8 & 1.0$\times$ & 0.19          & 18.0\% \\ 
    & CPU + VSAC\_QKeras                 & \textbf{86.3}  & 1018.3 & 1104.6 & 1.4$\times$  & \textbf{2.48} & 30.3\% & \textbf{68.2} & 330.9 & 399.1 & 1.0$\times$ & \textbf{0.14} & 38.4\% \\ 
    & CPU + VSAC\_MSQ                    & \textbf{86.6}  & 1019.6 & 1106.2 & 1.4$\times$  & \textbf{2.52} & 29.3\% & \textbf{68.3} & 330.5 & 398.8 & 1.0$\times$ & \textbf{0.15} & 36.1\% \\ 
    & CPU + VSAC\_APoT                   & \textbf{85.9}  & 1010.7 & 1096.7 & 1.4$\times$  & \textbf{2.48} & 30.3\% & \textbf{68.1} & 331.0 & 399.1 & 1.0$\times$ & \textbf{0.16} & 32.7\% \\ \hline

\parbox[t]{6mm}{\multirow{6}{*}{\rotatebox[origin=c]{90}{\shortstack{DeiT\_small \\ \cite{touvron2021training}}}}}
    & CPU                                & 1786.8          & 2015.4 & 3802.2 & 1.0$\times$ & 8.82          & 00.0\% & 300.1          & 653.0 & 953.1 & 1.0$\times$ & 0.60          & 00.0\% \\
    & CPU + VMAC~\cite{haris_secda_tflite_2023} & --              & --     & --     & --          & --            & --     & --             & --    & --    & --          & --            & --     \\ 
    & CPU + VMAC\_opt                    & 258.8           & 2015.4 & 2274.3 & 1.7$\times$ & 5.15          & 41.6\% & 148.2          & 653.0 & 801.2 & 1.2$\times$ & 0.42          & 30.2\% \\
    & CPU + VSAC\_QKeras                 & \textbf{229.9}  & 2004.8 & 2234.7 & 1.7$\times$ & \textbf{5.08} & 42.4\% & \textbf{143.4} & 653.0 & 796.3 & 1.2$\times$ & \textbf{0.29} & 51.0\% \\
    & CPU + VSAC\_MSQ                    & \textbf{230.1}  & 2004.8 & 2234.9 & 1.7$\times$ & \textbf{5.08} & 42.4\% & \textbf{143.9} & 654.2 & 798.1 & 1.2$\times$ & \textbf{0.31} & 48.3\% \\
    & CPU + VSAC\_APoT                   & \textbf{231.9}  & 2019.7 & 2251.6 & 1.7$\times$ & \textbf{5.08} & 42.4\% & \textbf{143.6} & 653.2 & 796.8 & 1.2$\times$ & \textbf{0.32} & 46.1\% \\ \hline

\end{tabular}
}
\end{table*}

\subsection{End-to-End Evaluation}

\subsubsection{Resource Utilization}

Table~\ref{tab:resource_utilization} presents the resource utilization of different accelerators (as individual blocks), including other accelerator blocks (e.g, DMAs), in the full hardware design (i.e., Vivado block design~\cite{amd_vivado_ipi_2019}) on both PYNQ-Z2 and Kria boards.
We targeted clock frequencies of 200 MHz for PYNQ-Z2 and 250 MHz for Kria.
These frequencies were chosen since they are the maximum achievable frequencies for the VMAC design on the respective boards.
Within our accelerator designs, we implemented PEs using LUTs instead of DSPs to perform a fair comparison between VMAC and VSAC.
When comparing the resource utilization of VMAC~\cite{haris_secda_tflite_2023} and VMAC\_opt on the PYNQ-Z2 platform, we observe that DSP usage is reduced by approximately $64\%$, along with a more efficient utilization of BRAM resources. 
Further comparison between VMAC\_opt and the VSAC designs reveals that the latter achieves, on average, a $20\%$ reduction in LUT utilization across both PYNQ-Z2 and Kria platforms, Table~\ref{tab:resource_utilization}. 

Figure~\ref{fig:resource-breakdown} illustrates the LUT utilization breakdown across different components of the VMAC\_opt and VSAC accelerators on both FPGA boards. 
Focusing on the \textit{compute} component, which primarily consists of the PE array, the VSAC designs, on average, achieve a $37\%$ reduction in LUT usage as compared to VMAC\_opt on the PYNQ-Z2 board, and a $53.8\%$ reduction on the Kria board. 
These savings can be attributed to the shift-PE design (Figure~\ref{fig:shift_pe_design}), which replaces the mult-PEs in each GEMM unit~\ref{fig:dma_weight_copy}(b).
Within the VSAC family of designs, VSAC\_MSQ and VSAC\_APoT exhibit higher LUT utilization than VSAC\_QKeras, due to the additional shift operations required for MSQ and APoT quantization schemes. 
Moreover, the difference in LUT consumption between VSAC\_MSQ and VSAC\_APoT stems from the intermediate product width ($ipw$) of the shift operations, Figure~\ref{fig:shift_pe_design}. 
Larger $ipw$ values correspond to larger shifters, which in turn demand more LUT resources.

\subsubsection{Accuracy}

Our proposed pipeline integrates both training and inference frameworks to enable the systematic evaluation of PoT-quantized DNN models on custom hardware accelerators.
\new{For the accuracy evaluation, the DNN models were trained and later converted using the \texttt{PoTAcc} pipeline (Section~\ref{sec:dnn_prep}).}
Note that, within our pipeline, the accuracy of a PoT-quantized DNN model can be evaluated on a given dataset using the inference framework(e.g., TFLite) using custom hardware accelerators.

\new{Table~\ref{tab:accuracy_test} reports the accuracy across different stages of the pipeline for the representative models on the CIFAR-10 and ImageNet datasets.
For CIFAR-10, MobileNetV2 (QKeras) achieves an accuracy degradation of $0.4\%$ after weight preprocessing, as compared to the trained model.
For ImageNet, the accuracy for ResNet18 (APoT) and MobileNetV2 (APoT) degrades by $1.0\%$ and $1.9\%$, respectively, after weight preprocessing, as compared to the trained models.
Such accuracy degradation can be attributed to the conversion of 32-bit floating-point activations to the $int8$ format during the model conversion stage.
More specifically, the \texttt{TFLite Converter} uses post-training quantization method to quantize the activations to the $int8$ format, which can lead to accuracy degradation, especially for models that are sensitive to quantization~\cite{jacob_2018_CVPR}.
Note that the weights are quantized to 4 bits during training, and their representation is changed from $pot\_float$ to $pot\_int^{e}$ during the model conversion and weight preprocessing stages, which results in minimal accuracy degradation (see Table~\ref{tab:weight_preprocessing_apot}).
When comparing the model conversion and weight preprocessing stages, the average accuracy degradation is $0.1\%$ across all models and datasets.
}
This result highlights the effectiveness of the DNN model conversion and weight preprocessing stages, confirming that custom accelerators can faithfully execute \new{a trained} PoT-quantized DNN model without experiencing significant accuracy degradation.

\subsubsection{Latency and Energy}

We evaluated the accelerators using synthetic DNNs (see Section~\ref{sec:dnn_prep}) on the ImageNet dataset~\cite{ILSVRC15}. 
In all experiments, the convolutional and FC layers of each DNN model run on the accelerators, while other layers of each DNN were executed on the CPU. 
Table~\ref{tab:end-to-end-evaluation} reports the end-to-end inference latency (ms/image) and energy consumption (J/image) for different DNNs on the PYNQ-Z2 and Kria boards. 
The CPU baselines were obtained using the maximum available cores on each board (2 cores for PYNQ-Z2 and 4 cores for Kria). 
Note that the rsults for VMAC~\cite{haris_secda_tflite_2023} on the Kria are omitted, as the design only targets PYNQ-Z2. 
Similarly, VMAC results for DeiT models are omitted since VMAC does not accelerate FC layers, which are the dominant layers in transformer models. 
For both boards, we execute $int8$ format PoT-quantized DNNs on the CPU, the CPU+VMAC, and the CPU+VMAC\_opt designs.
For the VSAC designs, we execute \new{$pot\_int^{e}$} format PoT-quantized DNNs on the CPU+VSAC\_QKeras, the CPU+VSAC\_MSQ, and the CPU+VSAC\_APoT designs.
Each experiment was repeated 100 times, and we report the average inference latency and energy consumption.

First, we compare the VMAC~\cite{haris_secda_tflite_2023} and VMAC\_opt designs on the PYNQ-Z2 board to evaluate the impact of our accelerator optimizations (see Section~\ref{sec:accelerator_design_optimization}). 
For convolutional layer-based DNNs (MobileNetV2, InceptionV1, ResNet18, and EfficientNet-L), VMAC\_opt achieves an average speedup of $1.3\times$ and an energy reduction of $20\%$ over VMAC. 
These gains result from the \new{weight-copy and DMA pre-load optimizations (see Section~\ref{sec:weight_copy_optimization} and Section~\ref{sec:dma_weight_preload_optimization})}, which reduce both the data transfer and computation time. 
MobileNetV2 benefits the least due to its relatively shallow architecture and smaller number of weights.

\new{Next, we compare the execution of PoT-quantized DNNs on the CPU and the CPU+VSAC on different hardware devices.}
On the PYNQ-Z2 board, CPU+VSAC, on average, achieves speedup of $2.3\times$ and an energy reduction of $52.6\%$ across all DNNs, while on the Kria board, on average, it achieves a speedup of $ 1.0\times$ and an energy reduction of $29.4\%$ compared to CPU-only execution.
Note that for KRIA, CPU+VSAC does not achieve any speedup in terms of $TotalTime$ because CPU execution outperforms by $1.7\times$ and $1.3\times$ for all accelerator designs for MobileNetV2 and EfficientNet-L, respectively.
\new{For MobileNetV2, approximately 50\% of the workload is offloaded to the accelerator (see ($T_{conv}+T_{fc}$) and ($T_{other}$) in Table V for the Kria) because our accelerator does not currently support layers for depthwise convolution.
Additionally, the early layers of MobileNetV2 have a smaller depth dimension, but larger output matrices. 
In our accelerator design, the activation dataflow is as follows: CPU $\rightarrow$ DMA $\rightarrow$ accelerator, and after computation, accelerator $\rightarrow$ DMA $\rightarrow$ CPU.
Such a dataflow increases $prep\_act$, $load\_act$, $send\_act$ and $store$ times, limiting the overall performance improvement of MobileNetV2.
Performance can be improved by offloading depthwise convolution layers and optimizing the activation dataflow. 
Specifically, storing accelerator outputs in the DMA buffer and directly reusing them for subsequent layers would reduce $load\_act$ and $store$ time.
The same optimization applies to EfficientNet-L, which also includes depthwise convolution layers currently executed on the CPU. 
Although the speedup is limited, EfficientNet-L still achieves significant energy reduction (20.6\% on average) due to shift-based execution and its deeper model architecture with more weights.
In general, further performance improvements on the Kria board would require offloading additional layer types and optimizing activation dataflow between consecutive layers. We plan to explore these optimizations in future work.
}
However, all CPU+VSAC designs achieve similar speedup and energy reduction across DNNs on the PYNQ-Z2, despite differences in shift-PE designs.
This outcome is expected, as all shift-PE designs process inputs in 1 clock cycle.
\new{On the Kria board, the energy reduction varies across CPU+VSAC designs because the accelerator uses twice as many shift-PEs as on the PYNQ-Z2 board. 
This amplifies the impact of shift-PE design complexity on resource utilization, leading to more noticeable differences in power consumption.}

For the comparison between CPU+VMAC\_opt and CPU+VSAC, we focus on the total convolutional and fc layer time, $T_{conv}+T_{fc}$, which represents the execution time of layers mapped to the accelerator (see Table~\ref{tab:end-to-end-evaluation}). 
On the PYNQ-Z2, the VSAC designs achieve an average speedup of $1.1\times$ across all DNNs compared to VMAC\_opt. 
This improvement is primarily due to the use of 4-bit weights in VSAC, in contrast to 8-bit weights in VMAC\_opt. 
The reduced weight precision effectively increases the local weight buffer (LWGT) capacity, thereby improving PE utilization. 
As illustrated in Figure~\ref{fig:obsrvation-vmac-accel}(a), a larger LWGT capacity correlates with higher accelerator performance, particularly for models such as InceptionV1 and ResNet18. 
In contrast, on the Kria board, the VSAC designs do not provide additional speedup over VMAC\_opt across the evaluated DNNs. 
Although the 8-GEMM configuration increases LWGT capacity, PE utilization does not improve further, resulting in comparable performance for both designs. 
With respect to total energy consumption, when comparing VSAC with to VMAC\_opt across all DNNs, both the PYNQ-Z2 and Kria boards achieved an average reduction of $4.9\%$ and $24.5\%$, respectively.
This reduction arises from the lower hardware cost of shift operations compared to multiplications. 
Although both complete in a single clock cycle, shift operations require fewer logic resources and hence consume less power compared to int8 multipliers.

When comparing the 4-GEMM configuration on the PYNQ-Z2 with the 8-GEMM configuration on the Kria for both VMAC\_opt and VSAC designs, and focusing on the $T_{conv}+T_{fc}$ time, the Kria achieves an average speedup of $1.5\times$ over the PYNQ-Z2.  
This improvement arises from doubling the number of GEMM units from 4 to 8, which effectively doubles the LWGT capacity and enhances PE utilization, thereby improving overall performance.

\section{Related Work}
\label{sec:related_work}


\subsection{PoT Quantization on DNNs}

While early researches~\cite{miyashita_convolutional_2016,gudovskiy_shiftcnn_2017,zhou_incremental_2017} introduced single-PoT-term logarithmic quantization
for convolutional layers in DNNs, subsequent studies explored multi-term and mixed quantization schemes to increase representational capacity.
APoT~\cite{li_additive_2020} employed additive PoT terms to extend the representation of a weight distribution using a quantization level, while MSQ~\cite{chang_mix_2021} combined double-term PoT and uniform quantization to optimize FPGA utilization.
RMSMP~\cite{chang_rmsmp_2021} refined MSQ by applying fine-grained, row-wise mixing of two types of quantizations, rather than layer-wise mixing.
More recently, CPoT~\cite{geng2024compact} smoothed multi-term PoT quantization levels both near zero and at higher magnitudes, to more accurately capture weight distributions.

Other studies focused on efficient training methodology.
DeepShift~\cite{elhoushi_deepshift_2021} enabled both left- and right-shift operations, S$^{3}$~\cite{li2021s3} decomposed weights into sign, sparsity, and shift components, and DenseShift~\cite{li2023densshift} removed zero from quantization levels to improve accuracy on large-scale datasets~\cite{krizhevsky2012imagenet, lin2014coco}.
Prezwlocka-Rus et al.~\cite{przewlocka-rus_power--two_2022} retained zero and studied pruning effects, while JLQ~\cite{jiang_jumping_2023} introduced non-uniform exponent “jumping” steps.
UPoT~\cite{xia_energy-and-area-efficient_2023} applied data-distribution-aware quantization, and PowerQuant~\cite{yvinec2023powerquant} enabled data-free quantization aligned with full-precision weight distributions.

More recent works extended PoT quantization beyond CNNs.
M2-ViT~\cite{liang2025m2vit} applied mixed PoT and uniform quantization to Vision Transformers, while ShifAddLLM~\cite{you2024shiftaddllm} and PoTPTQ~\cite{wang2025potptq} adapted PoT quantization for large language models (LLMs) to enable shift- and add-only inference.

These studies primarily focus on improving the accuracy of DNNs through their proposed PoT quantization methods, tailored to specific applications.
Among them several studies~\cite{gudovskiy_shiftcnn_2017,przewlocka-rus_power--two_2022, li2023densshift, jiang_jumping_2023,xia_energy-and-area-efficient_2023,chang_mix_2021,chang_rmsmp_2021}, have explored hardware design for shift-PE units or shift-accelerators based on FPGA/ASIC or CPU/GPU-based shift kernel implementations for their proposed PoT quantization method. 
Although these approaches demonstrate the feasibility of deploying PoT quantization on diverse hardware platforms, they do not fully address the challenges of end-to-end efficient inference for PoT-quantized DNNs.
In contrast, our work adopts three PoT quantization methods~\cite{qkeras,chang_mix_2021,li_additive_2020} and demonstrates efficient inference of PoT-quantized DNNs on FPGA-based accelerators integrated within an inference framework, enabling systematic evaluation of model accuracy, latency, and energy efficiency.


\subsection{Processing Element Design for PoT Quantization}

ShiftCNN~\cite{gudovskiy_shiftcnn_2017} proposed a shift-PE architecture based on right-shift operations and an adder for single-PoT-term quantization. 
Later, Vogel et al.~\cite{vogel_bit-shift-based_2019} introduced a left-shift-based PE design with 8-bit shift values for double-PoT-term quantization. 
Liu et al.~\cite{liu_optimize_2020} presented the BS-MUL architecture supporting 16-bit activations and 12-bit weights. 
Przewlocka-Rus et al.~\cite{przewlocka-rus_power--two_2022,przewlocka-rus_hwAccPoT_2023} investigated both single- and double-PoT-term shift-PE designs, incorporating a 4-bit shifter and a 16-bit accumulator. 
More recently, DenseShift~\cite{li2023densshift} and JLQ~\cite{jiang_jumping_2023} studied single-PoT-term shift-PE designs with modified quantization levels that exclude zero.
DenseShift~\cite{li2023densshift} also proposed a PE design that replaces shift operations with integer additions when activations are in FP16 format and weights are in PoT-quantized format, enabling efficient execution on CPUs.

Although these prior studies have introduced a variety of shift-PE designs tailored to specific PoT quantization methods, they do not explicitly address the optimization of intermediate product bit-width ($ipw$), which plays a critical role to determine the area and power efficiency of the shift-PE.


\subsection{Pipelines for PoT-quantized DNN Acceleration}

While there is a lack of end-to-end pipelines for PoT-quantized DNNs, several studies have proposed components that can be part of such a pipeline.
For instance, DeepShift~\cite{elhoushi_deepshift_2021} introduced CUDA kernels for shift-based inference on GPUs, while DenseShift~\cite{li2023densshift} developed CPU kernels supporting PoT-quantized weights and FP16 activations. 
UPoT~\cite{xia_energy-and-area-efficient_2023} designed an ASIC-based accelerator for PoT-quantized weights and activations, providing energy and area estimates. 

In the FPGA domain, a number of studies~\cite{gudovskiy_shiftcnn_2017,vogel_bit-shift-based_2019,liu_optimize_2020,nazari2023interlayer} proposed shift-based accelerators for specific PoT quantization methods.  
Similarly, Przewlocka-Rus et al.~\cite{przewlocka-rus_power--two_2022} trained models in PyTorch and converted them to ONNX~\cite{onnx} and TFLite formats to demonstrate model size reduction.
Their follow-up work~\cite{przewlocka-rus_hwAccPoT_2023} presented a MM accelerator to estimate energy consumption across different frequencies when mult-PEs are replaced with shift-PEs. 
However, these designs were not evaluated under end-to-end model execution.

In contrast, MSQ~\cite{chang_mix_2021}, FLIM-QNN~\cite{sun2022filmqnn}, and M2-ViT~\cite{liang2025m2vit} proposed FPGA-based accelerators for mixed quantization schemes (i.e., combining PoT with uniform quantization) and reported end-to-end latency. 
Nevertheless, these efforts did not describe how their PoT-quantized models were prepared for inference frameworks. 
The closest effort to our \texttt{PoTAcc} pipeline is MSQ~\cite{chang_mix_2021}, which trains DNNs using PyTorch~\cite{paszke2017pytorch} and maps them to FPGA hardware through TVM~\cite{chen2018tvm} and customized versions of the Versatile Tensor Accelerator (VTA)~\cite{moreau2019hardware}. 
While VTA is open-source, the accelerator design used in MSQ is not, and details on model preparation for TVM are not provided, limiting the applicability of their methodology to other PoT quantization methods.
To the best of our knowledge, no prior work has presented an open-source end-to-end pipeline for PoT-quantized DNN deployment.

\section{Conclusion}
\label{sec:conclusion}

In this paper, we have introduced \texttt{PoTAcc}, an open-source end-to-end pipeline for accelerating and evaluating PoT-quantized DNNs on resource-constrained edge devices, which integrates DNN model conversion, weight preprocessing, and deployment across CPU and custom accelerators.
\texttt{PoTAcc} supports pre-trained PoT-quantized DNN models from PyTorch and TensorFlow for TFLite execution, enabling efficient exploration of hardware designs.
\texttt{PoTAcc} includes an optimized mult-PE accelerator (\texttt{VMAC\_opt}) and three novel shift-PE-based \texttt{VSAC} accelerators targeting QKeras, MSQ, and APoT PoT quantization methods.

Our evaluation of FPGA platforms (PYNQ-Z2 and Kria) showed that \texttt{VSAC} reduced LUT utilization by $20\%$ on average compared to \texttt{VMAC\_opt}, \new{while not impacting accuracy (i.e., at most $0.4\%$ and $1.9\%$ degradation as compared to the baseline models on CIFAR-10 and ImageNet, respectively).}
Experiments on ImageNet CNN and Transformer models further showed that our proposed \texttt{VSAC} accelerators achieved up to $3.6\times$ speedup and $78.0\%$ energy reduction over CPU-only execution on the PYNQ-Z2 board, and a $1.0\times$ speedup with $29.4\%$ energy reduction on the Kria board. 
Overall, \texttt{PoTAcc} provides an open-source, flexible and extensible framework that advances the deployment of PoT-quantized DNNs for efficient AI inference on resource-constrained edge devices.

\new{Finally, it is important to note that without PoT-specific hardware, PoT-quantized DNN weights are represented in standard integer formats. To improve performance against integer-based execution on a CPU with more cores or a GPU, the shift-based accelerator needs to be further optimized by extending the support for more layer types and by optimizing the dataflow so that intermediate results remain on the accelerator. We plan to explore these improvements in future work.}


\balance
\bibliographystyle{IEEEtran}
\bibliography{bib}

\end{document}